\documentclass[preprint,12pt]{elsarticle}

\usepackage[utf8]{inputenc}
\usepackage[T1]{fontenc}
\usepackage{lmodern}
\usepackage{amsmath,amssymb}
\usepackage{booktabs}
\usepackage{array}
\usepackage{graphicx}
\usepackage{float}
\usepackage{multirow}
\usepackage{xcolor}
\usepackage{hyperref}
\usepackage{enumitem}
\usepackage{geometry}
\usepackage{caption}
\usepackage{longtable}
\usepackage{pdflscape}
\usepackage{placeins}
\usepackage{xspace}

\hypersetup{colorlinks=true, linkcolor=blue, citecolor=blue, urlcolor=blue}
\setlist[itemize]{leftmargin=*, itemsep=2pt, topsep=3pt}
\setlist[enumerate]{leftmargin=*, itemsep=2pt, topsep=3pt}

\newcolumntype{P}[1]{>{\raggedright\arraybackslash}p{#1}}

\journal{Chaos, Solitons \& Fractals}

\newcommand{\FEGPro}{FEG-Pro\xspace}
\newcommand{\GMAE}{\operatorname{GMAE}}
\newcommand{\FEDE}{\operatorname{FEDE}}

\newcommand{\code}[1]{\texttt{#1}}

\begin{document}

\begin{frontmatter}

\title{\FEGPro: Forecast-Error Growth Profiling for Finite-Horizon Instability Analysis of Nonlinear Time Series}

\author[psu]{Andrei Velichko\corref{cor1}}
\ead{velichkogf@gmail.com}
\author[tbd]{N'Gbo N'Gbo}
\author[unibz]{Bruno Carpentieri}
\author[balikesir,riphah]{Mudassir Shams}

\address[psu]{Institute of Physics and Technology, Petrozavodsk State University, Petrozavodsk, Russia}
\address[tbd]{School of Science and Engineering, International University of Grand Bassam, Grand Bassam, Cote d'Ivoire}
\address[unibz]{Faculty of Engineering, Free University of Bozen-Bolzano, Bolzano, Italy}
\address[balikesir]{Department of Mathematics, Faculty of Arts and Science, Balikesir University, Balikesir, Turkey}
\address[riphah]{Department of Mathematics and Statistics, Riphah International University, Islamabad, Pakistan}

\cortext[cor1]{Corresponding author.}

\begin{abstract}
Estimating the largest Lyapunov exponent from a scalar time series is difficult when the governing equations, tangent dynamics, and full state vector are unavailable. We propose \FEGPro, a forecast-error growth profiling framework for nonlinear scalar time series. The method constructs autocorrelation-guided sparse histories, performs distance-weighted k-nearest-neighbor multi-horizon forecasting, and analyzes the logarithmic growth of geometrically averaged forecast errors. Its primary output is the finite-horizon forecast-error growth slope, \(\lambda_{\mathrm{FEG}}\). When the error-growth curve supports a quasi-linear regime, this slope can be compared with reference largest Lyapunov exponents as an estimate of the dominant instability rate. The same pipeline also extracts the formal fit-selection regime, curvature, residual roughness after quadratic detrending, monotonicity, and forecast-error distribution entropy (FEDE) from signed multi-horizon errors. These secondary descriptors are intended not only as diagnostic controls for the slope, but also as candidate machine-learning features for nonlinear signal analysis, because they encode profile geometry and distributional uncertainty not captured by \(\lambda_{\mathrm{FEG}}\) alone. We evaluate the method on chaotic maps, Mackey--Glass delay dynamics, and scalar Lorenz-63 observables with known or reference exponents. Full-record experiments show good agreement in quasi-linear cases and meaningful curve-shape information in curved or weak profiles. A dyadic length-halving experiment on representative logistic, Mackey--Glass, and Lorenz records shows that residual roughness and mean FEDE often change monotonically and remain interpretable as record length decreases, even when the slope becomes biased or highly variable. The results support treating forecast-error growth as a structured profile and feature-generation framework rather than a single-number estimator.

\end{abstract}


\end{frontmatter}

\section{Introduction}

The largest Lyapunov exponent (LLE) is one of the central quantities used to characterize chaotic dynamics. It measures the mean exponential rate at which initially close trajectories separate and is therefore directly connected to predictability, instability, and the finite horizon over which a deterministic model remains useful. When the governing equations and tangent dynamics are known, the LLE can be computed by standard variational or Jacobian-based procedures. In experimental or archived data analysis, however, one often has access only to a scalar observable. The full state vector is unavailable, the equations may be unknown, and the measured signal may be a projection or nonlinear transformation of the underlying dynamics. This is the setting in which time-series methods for estimating Lyapunov exponents become necessary.

The classical route is phase-space reconstruction. Following the embedding viewpoint introduced by Takens, scalar observations are transformed into delay-coordinate vectors, and local divergence of nearby reconstructed trajectories is used as a proxy for instability \cite{takens1981,wolf1985,rosenstein1993,kantz1994,abarbanel1996,parlitz2016}. In this family, the methods of Wolf, Rosenstein, and Kantz remain influential. They usually produce a divergence curve, such as an average logarithmic neighbor distance versus time, and the LLE is inferred from a selected quasi-linear part of this curve. This approach is powerful, but it requires several choices that are not neutral in finite data: delay, embedding dimension, neighborhood definition, Theiler window, fitting range, filtering, and preprocessing. A large methodological literature is therefore devoted to delay and dimension selection, mutual information, false nearest neighbors, symbolic dynamics, and related criteria for stable reconstruction \cite{fraser1986,kennel1992,ma2006,matilla2021,tamma2016,zhu2016}. Noise and short records further complicate the problem, motivating robust variants and practical guidance on data length, filtering, and uncertainty of LLE estimates \cite{zeng1991,liu2005,yao2012,mehdizadeh2017,mehdizadeh2018,escot2023}. The important point for the present work is that these methods primarily target the exponent itself. The divergence curve is used mainly as an intermediate object from which a slope is extracted.

A second, closely related tradition is nonlinear forecasting of chaotic time series. Farmer and Sidorowich showed that short-term prediction can be learned from reconstructed states, and that the scaling of prediction errors is connected to dynamical invariants such as Lyapunov exponents and attractor dimension \cite{farmer1987}. Casdagli developed nonlinear prediction as a way to distinguish deterministic structure from randomness, and Sugihara and May demonstrated that nonlinear forecasting can separate chaos from measurement error by examining how predictive skill decays with prediction horizon \cite{casdagli1989,sugihara1990}. Later nonlinear time-series and empirical-dynamic-modeling frameworks continued to use local constant or local linear prediction, simplex projection, nearest-neighbor forecasting, and forecast skill as diagnostic tools \cite{hegger1999,kantzschreiber2004,sugihara1994,chang2017,bradley2015}. This literature is especially important because it already contains the key intuition behind our work: forecast errors are not merely nuisance quantities, but carry information about deterministic instability and predictability. Nevertheless, most studies use prediction error as a performance measure, a chaos-versus-noise diagnostic, or a route to a single instability estimate, rather than as a structured multi-coordinate profile.

A third relevant field is feature-based time-series analysis. Modern libraries such as hctsa, catch22, tsfresh, and theft extract hundreds or thousands of descriptors from a signal, including autocorrelation, distributional, spectral, entropy, and nonlinear features \cite{fulcher2017,lubba2019,christ2018,henderson2022}. Such descriptors have been used for classification, clustering, phenotyping, model selection, forecastability analysis, and meta-learning. In parallel, forecast residuals are widely used in hybrid models, residual correction, uncertainty quantification, conformal prediction, and forecast-combination frameworks \cite{deoliveira2021,ilhan2024,jensen2022,pessoa2025}. Entropy-based measures are also common descriptors of complexity and predictability \cite{shannon1948,guntu2020,papacharalampous2021}. However, these feature and residual frameworks usually optimize forecasting or classification performance. They do not usually connect residual features to a horizon-resolved forecast-error growth profile in which the slope, the formal fit-selection regime, curve geometry, and signed error-distribution entropy are extracted from the same multi-horizon forecast experiment.

The present study grows out of this intersection. In our previous work, a machine-learning forecast-error approach was proposed for estimating positive Lyapunov exponents in one-dimensional chaotic time series \cite{velichko2025chaos}. That method trained out-of-sample predictors, measured the exponential growth of geometrically averaged multi-horizon forecast errors, and inferred an exponent-related instability value from the error-growth slope. It was validated on canonical one-dimensional maps and showed that multi-horizon forecast errors can be used as a practical scalar-time-series route to positive Lyapunov exponents. Related follow-up studies by the same broader group applied similar forecast-error or kNN-LLE ideas to echo-state-network predictability, local Lyapunov proxies in neural systems, and stability analysis of numerical schemes \cite{belyaev2025esn,muruganantham2025local,shams2026}. These works support a broader view: prediction-error geometry can serve not only for forecasting, but also for dynamical characterization.

The present paper extends this line from an estimator of one number to a profiling framework. The central object is the multi-horizon forecast-error curve. If the logarithmic curve has a supported quasi-linear region, its slope is reported as \(\lambda_{\mathrm{FEG}}\), a finite-horizon forecast-error growth slope that can be interpreted in relation to the dominant Lyapunov exponent in controlled deterministic benchmarks. If the curve is curved, rough, nonmonotone, saturated, or locally unstable, the scalar slope alone is not enough. We therefore retain selected secondary descriptors: the formal fit-selection regime, curvature gain, residual roughness after quadratic detrending, monotonicity of error growth, and forecast-error distribution entropy (FEDE), computed from signed forecast errors at each horizon. These descriptors are not arbitrary decorations. Their choice reflects the physical and statistical structure of the forecast-error curve: curvature describes departure from a single exponential-like regime; roughness captures local irregularity after removing a smooth trend; monotonicity measures whether increasing horizon actually produces increasing error; and FEDE measures how the distribution of signed errors broadens or changes across the prediction horizon.

This profile-based formulation is particularly relevant for finite records. A short record may still preserve a meaningful forecast-error geometry even when the slope estimate is biased. Conversely, a plausible slope may be misleading when the error-growth curve is rough, curved, or supported by too few stable horizons. In the present benchmarks, this distinction becomes visible in a dyadic length-halving experiment. The primary slope remains the main endpoint, but residual roughness and mean FEDE show structured, often monotone changes as record length decreases. This observation suggests that secondary descriptors may become useful features for signal analysis in their own right. They can indicate how the error-growth profile degrades, and in future application studies they may provide information not captured by \(\lambda_{\mathrm{FEG}}\) alone.

The originality claimed here is therefore deliberately specific. We do not claim that forecast errors, nonlinear prediction, residual features, entropy, or Lyapunov estimation are individually new. Rather, the contribution is the unified construction of a multi-horizon forecast-error growth profile that combines a primary slope with curve-shape descriptors and signed forecast-error distribution entropy, all produced by the same scalar-time-series pipeline. To our knowledge, the existing literature does not combine these elements into a single framework with the same emphasis on finite-data profile stability.

A practical consequence is that \FEGPro should not be read as a replacement for classical LLE algorithms. It should be viewed as a reproducible forecast-error growth profiling procedure whose primary slope, \(\lambda_{\mathrm{FEG}}\), may approximate a dominant Lyapunov exponent only when the profile supports such an interpretation. At the same time, \FEGPro is a compact feature-extraction mechanism. In many signal-analysis tasks, including future machine-learning studies, the most useful information may not be the slope alone. Roughness can indicate local instability of the error-growth geometry, curvature can mark departure from one exponential-like regime, monotonicity can quantify whether longer horizons consistently increase error, and FEDE can describe distributional broadening of signed forecast errors. These quantities are physically motivated by the prediction process and statistically useful because they summarize distinct aspects of finite-data predictability. The length-halving experiment reported below provides an initial consistency check: several of these descriptors respond monotonically or compactly when record length is reduced, which supports their interpretation as candidate features rather than arbitrary by-products.

The manuscript is organized as follows. Section~2 summarizes the conceptual contribution. Section~3 defines the \FEGPro pipeline, including autocorrelation-guided sparse histories, kNN multi-horizon prediction, the logarithmic GMAE curve, fit selection, diagnostic descriptors, and FEDE. Section~4 describes the compact diagnostic profile. Section~5 presents the numerical benchmarks, including the benchmark systems, reference values, universal configuration, and full-record results for discrete maps, Mackey--Glass delay dynamics, and Lorenz-63 scalar observables. Section~6 reports the dyadic length-halving experiment and shows how the primary slope and selected secondary descriptors behave as record length decreases. The final sections discuss the implications, limitations, and future use of the profile descriptors as features for nonlinear signal analysis.

\section{Conceptual contribution}

Against this literature background, the contribution of \FEGPro can be summarized by five points.

\begin{enumerate}
    \item \textbf{Forecast-error growth as finite-horizon instability proxy.} If nearby reconstructed states become harder to predict as horizon increases, the multi-horizon forecast-error curve contains information about dynamical instability.
    \item \textbf{Autocorrelation-guided reconstruction.} The history span and the number of sparse historical samples are derived from empirical autocorrelation crossings rather than independently tuned for each system.
    \item \textbf{Profile rather than scalar-only output.} The slope \(\lambda_{\mathrm{FEG}}\) is the main finite-horizon forecast-error growth endpoint, but the selected fit regime, curvature, roughness, monotonicity, and FEDE are reported as interpretable coordinates of the same forecast-error profile.
    \item \textbf{Feature-oriented signal description.} The secondary descriptors are retained as a compact feature vector for downstream nonlinear signal analysis. They can be used by machine-learning pipelines, statistical classifiers, anomaly detectors, or comparative studies even when the exact asymptotic Lyapunov exponent is unavailable.
    \item \textbf{Robustness as profile stability.} Under record shortening, the question is not only whether the slope remains accurate; it is also whether the profile descriptors indicate the loss of reliable error-growth geometry.
\end{enumerate}

The method is therefore best regarded as a feature-generation framework for nonlinear predictability. One generated feature is \(\lambda_{\mathrm{FEG}}\). Other features become important when the ideal single-exponential picture is only partially supported. The selected descriptors are intentionally compact rather than exhaustive: they encode the main ways in which a forecast-error curve departs from the ideal single-slope picture while remaining simple enough to be exported as interpretable signal features.

\section{Forecast-error growth pipeline}

Let a scalar time series be
\begin{equation}
    x_0,x_1,\ldots,x_{N-1}, \qquad x_i=x(t_i).
\end{equation}
For a discrete map, \(t_i=i\). For a continuous-time system sampled at constant interval \(\Delta t\), the physical prediction horizon associated with an integer horizon \(h\) is \(\tau_h=h\Delta t\).

\subsection{Autocorrelation scales}

For lag \(k\), the empirical autocorrelation is computed on overlapping segments:
\begin{equation}
    C(k)=\operatorname{corr}\left( x_0,\ldots,x_{N-k-1}; x_k,\ldots,x_{N-1}\right).
\end{equation}
Two crossing indices are extracted:
\begin{align}
    Z_{+-} &= \text{first positive-to-negative crossing index},\\
    Z_{-+} &= \text{next negative-to-positive crossing index}.
\end{align}
These indices are empirical scales of the scalar observable. They are not interpreted as universal dynamical invariants; they are used to define a data-driven historical representation and horizon grid.

Operationally, \(Z_{+-}\) marks the first lag at which the observable loses positive self-similarity and becomes anticorrelated, while \(Z_{-+}\) marks the following return from negative to positive correlation. Their sum is used as a practical memory scale for the measured scalar signal. The history span is chosen as a multiple of this scale, and the sparse historical coordinates are spread across the resulting window. Thus the method does not search independently over a fixed delay and embedding dimension for every system; it constructs a reproducible history representation from the autocorrelation structure of the observed series.

The same scale also guides the optional internal resampling used in the common configuration. When \(Z_{+-}\) is very large, the record can contain redundant oversampling relative to the predictive scale. The adaptive compression step reduces this redundancy while preserving a minimum number of samples and the full time span. This step should be understood as an operational preprocessing rule for comparable forecast-error profiling, not as a lossless transformation or a dynamical invariant. In practical terms, the intention is to reduce excessive temporal redundancy before the kNN search, while keeping the autocorrelation-defined predictive scale, the chronological order, and the physical time span of the record explicit. The same rule is applied reproducibly across the benchmark files, so the resampling step is part of the reported operational configuration rather than a per-series optimization.

\subsection{Sparse historical vectors}

The history span is
\begin{equation}
    H = \operatorname{round}\left( \alpha_H (Z_{+-}+Z_{-+}) \right),
\end{equation}
with \(\alpha_H=3\) in the current configuration. The number of historical samples is bounded by
\begin{equation}
    m = \min\{m_{\max},\max(m_{\min},Z_{+-})\},
\end{equation}
where \(m_{\min}=3\) and \(m_{\max}=32\). The offsets are spread across the history window:
\begin{equation}
    \ell_j=\operatorname{round}\left(\frac{j}{m-1}(H-1)\right), \qquad j=0,\ldots,m-1.
\end{equation}
The historical vector at base index \(i\) is
\begin{equation}
    \mathbf{s}_i = \left(x_{i+\ell_0},x_{i+\ell_1},\ldots,x_{i+\ell_{m-1}}\right),
\end{equation}
where the indices span a window of length \(H\). This is a sparse history-window representation rather than a conventional fixed-delay embedding.

\subsection{Multi-horizon kNN forecasting}

For each prediction horizon \(h\), the target is
\begin{equation}
    y_i^{(h)}=x_{i+H+h-1}.
\end{equation}
The historical vectors are split chronologically into a training part and a testing part. Features are standardized using only the training statistics. A distance-weighted \(K\)-nearest-neighbor regressor gives
\begin{equation}
    \hat{x}_{i+h} = F_h(\mathbf{s}_i),
\end{equation}
with \(K=3\) in the reported experiments. The signed forecast error is
\begin{equation}
    \eta_i(h)=x_{i+H+h-1}-\hat{x}_{i+h}.
\end{equation}

\subsection{Forecast-error curve}

For each horizon, the absolute errors are summarized by the geometric mean absolute error:
\begin{equation}
    \GMAE(h)=\exp\left[\frac{1}{n_h}\sum_i \log\left(|\eta_i(h)|+\epsilon\right)\right].
\end{equation}
The forecast-error growth curve is
\begin{equation}
    g(h)=\log \GMAE(h).
\end{equation}
For continuous-time data, fitting can be performed versus the physical horizon \(\tau_h=h\Delta t\); for discrete maps it is performed per iteration. If \(g(h)\) is locally linear, then
\begin{equation}
    g(h)\approx a+\lambda_{\mathrm{FEG}}\tau_h,
\end{equation}
so that the slope \(\lambda_{\mathrm{FEG}}\) is the finite-horizon forecast-error growth slope. When this local linear regime is supported by the diagnostics, \(\lambda_{\mathrm{FEG}}\) can be compared with a largest Lyapunov exponent; otherwise it should be read as a finite-horizon forecast-instability descriptor.

\subsection{Horizon grid}

The horizon grid always includes a dense prefix,
\begin{equation}
    h=1,2,\ldots,h_{\mathrm{dense}},
\end{equation}
with \(h_{\mathrm{dense}}=15\) in the current configuration. If \(Z_{+-}\) is larger than the small-scale threshold, an additional autocorrelation-based grid is added between \(Z_{+-}+1\) and \((Z_{+-}+1)M_h\), where \(M_h=15\). The maximum horizon is capped by a fraction of the available series length. This design keeps the early fast-growth region for maps while still sampling longer horizons in delayed and continuous-time systems.

\subsection{Candidate fits and deterministic definition of \texorpdfstring{\(\lambda_{\mathrm{FEG}}\)}{lambdaFEG}}

The reported slope is not assigned by a qualitative label. It is defined by a deterministic selection rule applied to the forecast-error curve. Let
\begin{equation}
    \mathcal{C}=\{(x_i,y_i)\}_{i=1}^{M},
    \qquad y_i=\log \GMAE(h_i),
\end{equation}
where \(x_i\) is either the physical horizon time or the horizon step, depending on the selected fitting axis. The algorithm fits four candidate descriptions of \(\mathcal{C}\): a single line, early-window lines over the first \(k=2,\ldots,10\) points, a two-line piecewise model, and a quadratic curve. Only three regimes can define the reported \(\lambda_{\mathrm{FEG}}\): \code{early\_linear\_window}, \code{two\_lines\_first\_slope}, and \code{single\_line}. The quadratic fit is used only for shape descriptors.

\begin{table}[H]
\centering
\caption{Algorithm 1. Deterministic definition of \(\lambda_{\mathrm{FEG}}\) and \code{selected\_fit}. The thresholds correspond to the common configuration used in the experiments.}
\label{tab:algorithm}
\label{alg:lambda-feg}
\scriptsize
\begin{tabular}{P{0.17\textwidth}P{0.50\textwidth}P{0.25\textwidth}}
\toprule
Stage & Rule & Output \\
\midrule
Input curve &
\(\mathcal{C}=\{(x_i,y_i)\}_{i=1}^{M}\), where \(y_i=\log\GMAE(h_i)\). The fitting coordinate \(x_i\) is either horizon time or horizon step. &
-- \\
\addlinespace[2pt]
Candidate fits &
\(L_{\rm single}(x)=a_{\rm single}x+b_{\rm single}\); \quad
\(L_{{\rm early},k}(x)=a_{{\rm early},k}x+b_{{\rm early},k}\), \(k=2,\ldots,10\); \quad
\(L_{\rm two}(x)=a_Lx+b_L\) for \(x<x_s\) and \(a_Rx+b_R\) for \(x\ge x_s\); \quad
\(Q(x)=c_2x^2+c_1x+c_0\). &
Candidate slopes and shape fits \\
\addlinespace[2pt]
Early regime &
Allowed only when \(Z_{pm}^{\rm used}\le 15\). For each \(k=2,\ldots,10\), fit the first \(k\) points and compute \(R_k^2\) and
\(\mathrm{nMAE}_k=\mathrm{MAE}_k/(\max y-\min y+\epsilon)\). Keep windows satisfying
\(R_k^2\ge0.98\) and \(\mathrm{nMAE}_k\le0.08\). If at least one window is accepted, select
\[
    k^*=\arg\max_k \left(R_k^2,-\mathrm{nMAE}_k,k\right).
\]
Thus the priority is maximum \(R^2\), then minimum normalized MAE, then more points. &
\(\lambda_{\mathrm{FEG}}=a_{{\rm early},k^*}\), \quad
\code{selected\_fit=early\_linear\_window} \\
\addlinespace[2pt]
Two-line regime &
Used only if the early regime is not selected. The two-line model must be the best candidate by \(R^2\) and must satisfy
\[
R^2_{\rm two}\ge0.98, \quad
R^2_{{\rm two},\min}\ge0.95, \quad
\frac{R^2_{\rm two}-R^2_{\rm single}}{|R^2_{\rm single}|+\epsilon}\ge0.03 .
\]
Each segment must contain at least five points. &
\(\lambda_{\mathrm{FEG}}=a_L\), \quad
\code{selected\_fit=two\_lines\_first\_slope} \\
\addlinespace[2pt]
Single-line regime &
If neither the early regime nor the two-line regime is selected, use the line fitted to the full forecast-error curve. &
\(\lambda_{\mathrm{FEG}}=a_{\rm single}\), \quad
\code{selected\_fit=single\_line} \\
\addlinespace[2pt]
Quadratic role &
The quadratic fit may have the best \(R^2\), but it does not replace \(\lambda_{\mathrm{FEG}}\). It is used to quantify curvature and residual roughness. &
\(D_{\rm curv}\), \(D_{\rm rough}\) \\
\bottomrule
\end{tabular}
\end{table}

Categorical labels stored by the implementation, such as \code{clean\_linear}, \code{curved\_growth}, or \code{weak\_or\_mixed}, are retained in the JSON archive for inspection. They are not used as primary endpoints in this manuscript. The formal trace reported in the main tables is \code{selected\_fit}, because it specifies which of the three regimes actually defined \(\lambda_{\mathrm{FEG}}\).

The interpretation of Table~\ref{tab:algorithm} is therefore hierarchical. If the selected regime is \code{early\_linear\_window}, \code{two\_lines\_first\_slope}, or \code{single\_line} with strong line-fit support, \(\lambda_{\mathrm{FEG}}\) is read as a finite-horizon slope of the forecast-error growth curve and may be compared with a reference largest Lyapunov exponent when such a reference is available. If the profile is best described as curved, the quadratic fit still does not define \(\lambda_{\mathrm{FEG}}\). In that case the reported slope is retained only as a finite-horizon summary, and the curvature and roughness descriptors become essential for interpretation. If the profile is weak, mixed, rough, or supported by too few stable horizons, \(\lambda_{\mathrm{FEG}}\) should not be over-interpreted as an exponent.

\section{Compact diagnostic profile}

\begin{figure}[t]
    \centering
    \includegraphics[width=0.85\textwidth]{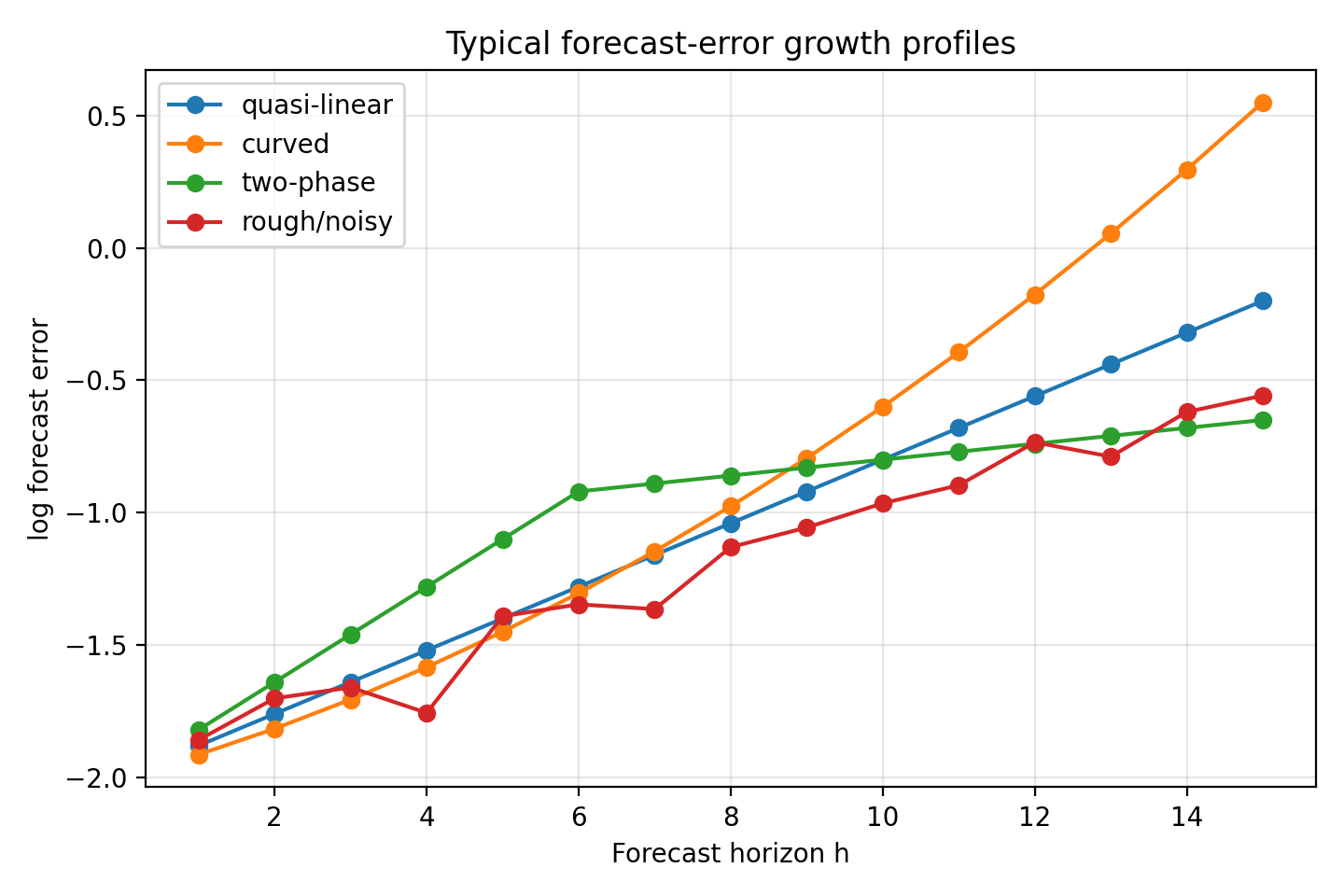}
    \caption{Typical forecast-error growth profiles considered by \FEGPro. A clean quasi-linear region supports comparison of the FEG slope with a reference Lyapunov exponent. Curved, two-phase, rough, and noisy profiles require additional diagnostics.}
    \label{fig:profile-types}
\end{figure}

The same forecast-error curve \(g(h)\) is used to define secondary descriptors. They are not separate models of the time series; they are coordinates of the forecast-error profile.

\subsection{Linearity and fit-selection support}

The single-line coefficient of determination \(R^2_{\mathrm{single}}\) and the formal \code{selected\_fit} regime summarize the line-fit support behind the reported slope. A high numerical slope is not sufficient by itself: the same value has a different interpretation when it comes from a short early window, from a first segment of a two-line fit, or from a single line over the full horizon range. For this reason, the main text reports formal fit quantities and curve-shape descriptors rather than qualitative confidence labels.

\subsection{Curvature}

The quadratic fit is used to measure departure from a single straight line. In the current implementation the curvature gain is
\begin{equation}
    D_{\mathrm{curv}}=
    \frac{\mathrm{MAE}_{\mathrm{single}}-\mathrm{MAE}_{\mathrm{quad}}}
    {\mathrm{MAE}_{\mathrm{single}}+\epsilon}.
\end{equation}
A large positive value means that a quadratic curve explains the forecast-error profile substantially better than one line.

\subsection{Residual roughness}

Let
\begin{equation}
    q(h)=c_0+c_1\tau_h+c_2\tau_h^2
\end{equation}
be the fitted quadratic trend and
\begin{equation}
    r(h)=g(h)-q(h)
\end{equation}
the residual profile. The main roughness descriptor used in the robustness analysis is
\begin{equation}
    D_{\mathrm{rough}}=
    \frac{\operatorname{mean}\left(|\Delta^2 r(h)|\right)}{\operatorname{range}(g)+\epsilon}.
\end{equation}
This quantity differs from residual amplitude. It measures local irregularity after removal of a smooth quadratic component.

\subsection{Monotonicity}

The monotonicity descriptor is
\begin{equation}
    D_{\mathrm{mono}}=
    \frac{1}{H_f-1}\sum_i \mathbf{1}\{g(h_{i+1})>g(h_i)\}.
\end{equation}
Values near one indicate an almost always increasing error-growth curve. Lower values indicate saturation, reversals, or mixed local behavior.

\subsection{Forecast-error distribution entropy}

FEDE analyzes the signed forecast-error distribution directly. For reproducibility, the histogram bin edges are fixed separately for each analyzed record or fragment. They are computed from the pooled signed forecast errors over all horizons, using the global minimum and maximum of this pooled set and \(B\) equal-width bins. The same bin edges are then used to estimate \(p_b(h)\) for every horizon \(h\). Let \(p_b(h)\), \(b=1,\ldots,B\), be the corresponding empirical probabilities. In the article, \(\FEDE(h)\) denotes the normalized entropy
\begin{equation}
    \FEDE(h)=
    \frac{-\sum_{b=1}^B p_b(h)\log p_b(h)}{\log B}.
\end{equation}
Thus \(0\leq \FEDE(h)\leq 1\). The reported scalar summaries are
\begin{align}
    \overline{\FEDE} &= \frac{1}{H_f}\sum_{h=1}^{H_f}\FEDE(h),\\
    \Delta \FEDE &= \FEDE(H_f)-\FEDE(1),\\
    s_{\FEDE}^{\mathrm{early}} &= \operatorname{slope}\left(\FEDE(h), h=1,\ldots,h_e\right).
\end{align}
In the experiments below, \(B=32\) and \(h_e=5\). FEDE complements the logarithmic mean error curve: \(g(h)\) measures the scale of forecast errors, whereas FEDE measures distributional uncertainty.

\begin{table}[t]
\centering
\caption{Compact \FEGPro profile used in the robustness analysis.}
\label{tab:compact-profile}
\small
\begin{tabular}{P{0.22\textwidth}P{0.23\textwidth}P{0.45\textwidth}}
\toprule
Group & Article symbol & Interpretation \\
\midrule
Primary slope & \(\lambda_{\mathrm{FEG}}\) & Finite-horizon FEG slope; comparison with a Lyapunov exponent only when the selected region is supported. \\
Fit-selection support & \(R^2_{\rm single}\), \code{selected\_fit} & Formal line-fit support and the regime used to define the selected slope. \\
Curvature & \(D_{\rm curv}\) & Improvement of quadratic profile over a single line. \\
Roughness & \(D_{\rm rough}\) & Local irregularity after quadratic detrending. \\
Monotonicity & \(D_{\rm mono}\) & Fraction of increasing steps in the error-growth curve. \\
Entropy & \(\overline{\FEDE},\Delta\FEDE\) & Distributional uncertainty of signed forecast errors. \\
\bottomrule
\end{tabular}
\end{table}

\FloatBarrier

\section{Numerical benchmarks}

\subsection{Benchmark systems and reference values}

This section states explicitly which systems were used to validate the pipeline. The purpose is to test one common scalar-observable procedure across qualitatively different deterministic systems: memoryless maps, a scalar delay system, and a continuous-time three-dimensional flow.

\subsubsection{One-dimensional discrete chaotic maps}

For maps, the data consist of iterates
\begin{equation}
    x_{n+1}=f(x_n), \qquad n=0,1,2,\ldots,
\end{equation}
stored in the common \code{t,module} CSV format with \(t=n\). Initial transients were removed before saving the benchmark segments. The typical saved length is \(N=20000\) samples.

The logistic map is
\begin{equation}
    x_{n+1}=r x_n(1-x_n), \qquad r=4,
\end{equation}
with reference exponent \(\lambda=\ln 2\).

The skew tent map is
\begin{equation}
    x_{n+1}=T_p(x_n)=
    \begin{cases}
        x_n/p, & 0\leq x_n<p,\\
        (1-x_n)/(1-p), & p\leq x_n\leq 1,
    \end{cases}
\end{equation}
with reference exponent
\begin{equation}
    \lambda=-p\ln p-(1-p)\ln(1-p).
\end{equation}

The beta map is
\begin{equation}
    x_{n+1}=\beta x_n \bmod 1,
\end{equation}
whose derivative magnitude is \(\beta\) almost everywhere, giving reference exponent \(\lambda=\ln\beta\).

The Chebyshev map of degree \(d\) is
\begin{equation}
    x_{n+1}=\cos\left(d\arccos x_n\right), \qquad x_n\in[-1,1],
\end{equation}
with reference exponent \(\lambda=\ln d\). High-degree Chebyshev and beta-map cases are deliberately difficult because their error-growth curves can saturate after only a few horizons.

\begin{table}[t]
\centering
\caption{Discrete-map benchmark families and reference exponents.}
\label{tab:map-definitions}
\small
\begin{tabular}{P{0.17\textwidth}P{0.26\textwidth}P{0.25\textwidth}P{0.22\textwidth}}
\toprule
Family & Tested parameters & Reference exponent & Figure label \\
\midrule
Logistic & \(r=4\) & \(\ln 2\) & Log r=4 \\
Skew tent & \(p=0.15,0.25,0.35,0.45\) & \(-p\ln p-(1-p)\ln(1-p)\) & ST p \\
Beta map & \(\beta=1.2,1.5,1.8,2.2,2.8,3.3,4.5\) & \(\ln\beta\) & B value \\
Chebyshev & \(d=2,3,4,5,6,8,10\) & \(\ln d\) & T degree \\
\bottomrule
\end{tabular}
\end{table}

\subsubsection{Mackey-Glass delay dynamics}

The Mackey-Glass equation \cite{mackey1977} is a scalar delay-differential system:
\begin{equation}
    \frac{dx(t)}{dt}=
    \frac{\beta x(t-\tau)}{1+x(t-\tau)^n}-\gamma x(t).
\end{equation}
The benchmark uses the standard parameters
\begin{equation}
    \beta=0.2,\qquad \gamma=0.1,\qquad n=10,
\end{equation}
with delays
\begin{equation}
    \tau\in\{10,12,14,16,17,18,20,23,30,50\}.
\end{equation}
The scalar variable \(x(t)\) is saved as \code{module}. After transient removal, each file contains approximately \(24430\) samples over about \(6000\) time units. This benchmark is important because the effective state depends on delayed history; the sparse historical representation is therefore not merely a convenience but part of the reconstruction.

\subsubsection{Lorenz-63 scalar observable}

The Lorenz-63 system \cite{lorenz1963} is
\begin{align}
    \dot{x} &= \sigma(y-x),\\
    \dot{y} &= x(\rho-z)-y,\\
    \dot{z} &= xy-\beta z.
\end{align}
The benchmark uses
\begin{equation}
    \sigma=10,\qquad \beta=\frac{8}{3},
\end{equation}
with \(\rho=24.0,24.5,\ldots,30.0\). Trajectories are integrated with \(\Delta t=0.01\), a total time of \(1100\), and transient removal of the first \(100\) time units, leaving approximately \(100000\) samples. The analyzed scalar observable is the module
\begin{equation}
    r(t)=\sqrt{x(t)^2+y(t)^2+z(t)^2},
\end{equation}
which is saved as \code{t,module}. Reference values are largest Lyapunov exponents of the full Lorenz flow obtained from the variational equations with QR orthonormalization. The comparison is therefore strict: \FEGPro observes only a scalar projection, not the full tangent dynamics.

\subsubsection{Universal configuration}

All full-record benchmarks and length-halving fragments were processed with one common configuration. The main settings are summarized in Table~\ref{tab:universal-config}. The aim is not to claim that these values are globally optimal, but to test whether a reproducible physics-guided default works across heterogeneous series.

\begin{table}[t]
\centering
\caption{Current universal configuration.}
\label{tab:universal-config}
\small
\begin{tabular}{P{0.30\textwidth}P{0.58\textwidth}}
\toprule
Component & Setting \\
\midrule
CSV input & Columns \code{t,module}. \\
Internal resampling & Adaptive FFT compression only if \(Z_{+-}>25\); minimum resampled length 256. \\
History span & \(H=\operatorname{round}(3(Z_{+-}+Z_{-+}))\), clipped to available length. \\
Historical dimension & \(m=\operatorname{clamp}(Z_{+-},3,32)\). \\
kNN model & Distance-weighted \(K=3\), chronological train/test split 60/40, standardized features. \\
Horizon grid & Dense prefix \(1,\ldots,15\); optional ACF grid to \(15(Z_{+-}+1)\); maximum horizon 30\% of series length. \\
Fit selection & Early-window rule for small-\(Z_{+-}\) cases; otherwise single-line/two-line/quadratic diagnostics with conservative selection. \\
FEDE & Histogram entropy of signed forecast errors; \(B=32\); normalized by \(\log B\); early FEDE slope over first five horizon points. \\
\bottomrule
\end{tabular}
\end{table}

\FloatBarrier

\subsection{Full-record benchmark results}

\subsubsection{Discrete-map benchmark}

For the one-dimensional maps, the method operates per iteration. In all map files, the adaptive resampling block did not modify the series because the autocorrelation crossing scales were below the resampling threshold. The dense early horizon prefix was therefore essential: the informative growth interval can be very short before saturation.

\begin{figure}[t]
    \centering
    \includegraphics[width=0.95\textwidth]{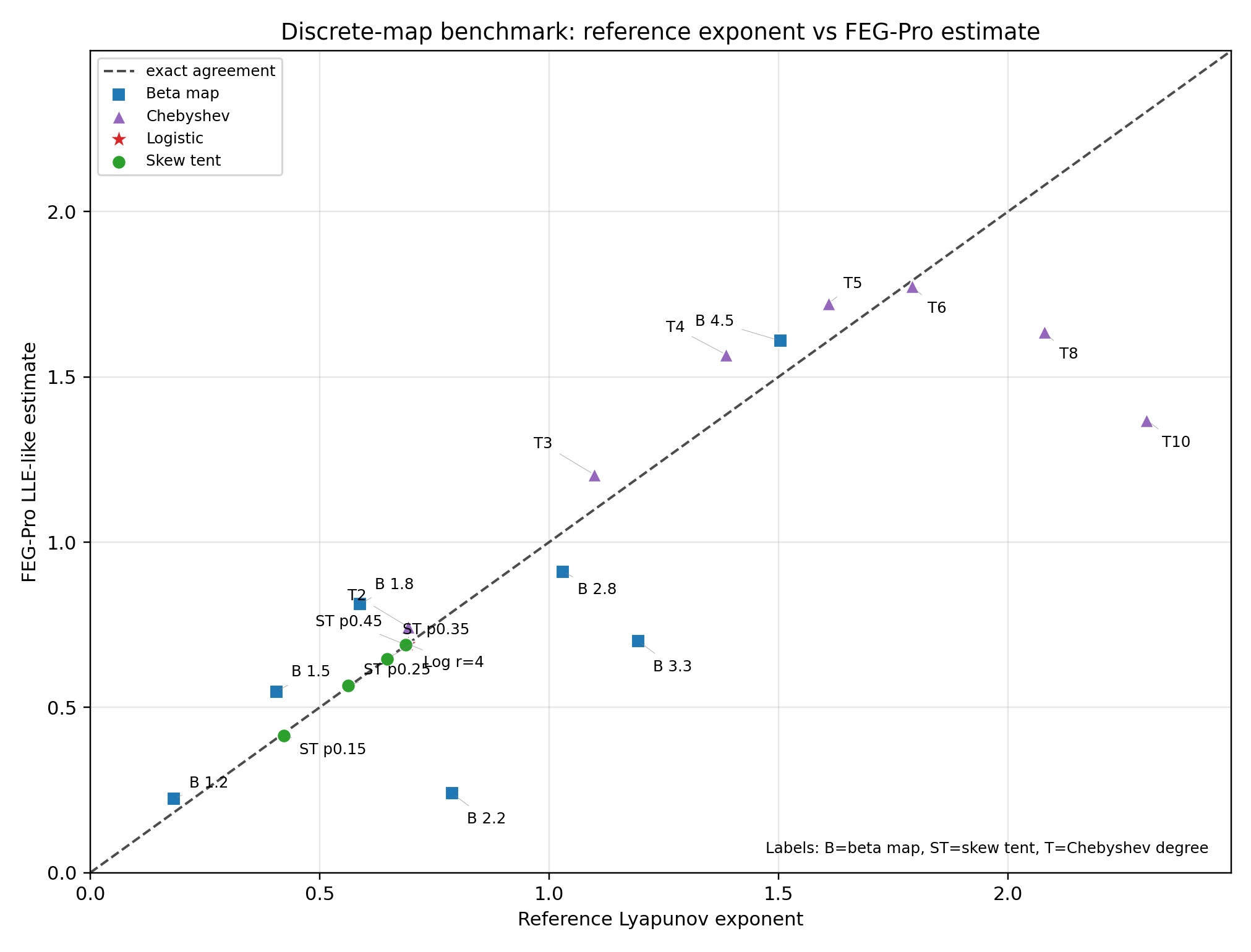}
    \caption{Reference Lyapunov exponent versus \FEGPro \(\lambda_{\mathrm{FEG}}\) for the discrete-map benchmark set. The diagonal indicates exact agreement. Point labels identify each map: Log is the logistic map, ST is the skew tent map, B is the beta map, and T is the Chebyshev degree. The largest deviations occur in beta-map and high-degree Chebyshev cases where the error-growth profile is less suitable for a single early slope.}
    \label{fig:discrete-scatter}
\end{figure}

Figure~\ref{fig:discrete-scatter} compares reference exponents and \FEGPro estimates. The logistic and skew-tent cases lie close to the diagonal. Several beta-map and high-degree Chebyshev cases deviate more strongly. These are not random failures: they are cases where the forecast-error curve often saturates quickly or is less suitable for a single early slope. Table~\ref{tab:discrete-detailed} therefore reports the formal \code{selected\_fit} regime, rather than qualitative confidence labels.

\begin{table}[t]
\centering
\caption{Detailed discrete-map results. The code column matches the labels in Fig.~\ref{fig:discrete-scatter}. The final column gives the formal slope-selection regime used to define \(\lambda_{\mathrm{FEG}}\).}
\label{tab:discrete-detailed}
\scriptsize
\begin{tabular}{llcrrrl}
\toprule
Code & Family & Parameter & Reference & Estimate & Abs. error & \code{selected\_fit} \\
\midrule
Log4 & Logistic & $r=4$ & 0.6931 & 0.6916 & 0.0015 & \code{early\_linear\_window} \\
ST0.15 & Skew tent & $p=0.15$ & 0.4227 & 0.4156 & 0.0071 & \code{early\_linear\_window} \\
ST0.25 & Skew tent & $p=0.25$ & 0.5623 & 0.5665 & 0.0042 & \code{early\_linear\_window} \\
ST0.35 & Skew tent & $p=0.35$ & 0.6474 & 0.6474 & 0.0000 & \code{early\_linear\_window} \\
ST0.45 & Skew tent & $p=0.45$ & 0.6881 & 0.6893 & 0.0011 & \code{early\_linear\_window} \\
B1.2 & Beta map & $\beta=1.2$ & 0.1823 & 0.2238 & 0.0415 & \code{early\_linear\_window} \\
B1.5 & Beta map & $\beta=1.5$ & 0.4055 & 0.5478 & 0.1424 & \code{early\_linear\_window} \\
B1.8 & Beta map & $\beta=1.8$ & 0.5878 & 0.8128 & 0.2250 & \code{early\_linear\_window} \\
B2.2 & Beta map & $\beta=2.2$ & 0.7885 & 0.2403 & 0.5482 & \code{early\_linear\_window} \\
B2.8 & Beta map & $\beta=2.8$ & 1.0296 & 0.9113 & 0.1183 & \code{early\_linear\_window} \\
B3.3 & Beta map & $\beta=3.3$ & 1.1939 & 0.7016 & 0.4923 & \code{early\_linear\_window} \\
B4.5 & Beta map & $\beta=4.5$ & 1.5041 & 1.6093 & 0.1053 & \code{early\_linear\_window} \\
T2 & Chebyshev & $d=2$ & 0.6931 & 0.7437 & 0.0506 & \code{early\_linear\_window} \\
T3 & Chebyshev & $d=3$ & 1.0986 & 1.2044 & 0.1057 & \code{early\_linear\_window} \\
T4 & Chebyshev & $d=4$ & 1.3863 & 1.5673 & 0.1810 & \code{early\_linear\_window} \\
T5 & Chebyshev & $d=5$ & 1.6094 & 1.7227 & 0.1132 & \code{early\_linear\_window} \\
T6 & Chebyshev & $d=6$ & 1.7918 & 1.7738 & 0.0180 & \code{early\_linear\_window} \\
T8 & Chebyshev & $d=8$ & 2.0794 & 1.6353 & 0.4441 & \code{early\_linear\_window} \\
T10 & Chebyshev & $d=10$ & 2.3026 & 1.3690 & 0.9336 & \code{early\_linear\_window} \\
\bottomrule
\end{tabular}
\end{table}

The family-level errors are consistent with the profile interpretation. Logistic and skew-tent maps are recovered accurately. Beta maps and Chebyshev maps contain more difficult fast-expansion cases. In all discrete-map cases the reported slope is selected by the early-window regime, which is appropriate for short fast-growth intervals followed by saturation.

\subsubsection{Mackey-Glass benchmark}

\begin{figure}[t]
    \centering
    \includegraphics[width=0.82\textwidth]{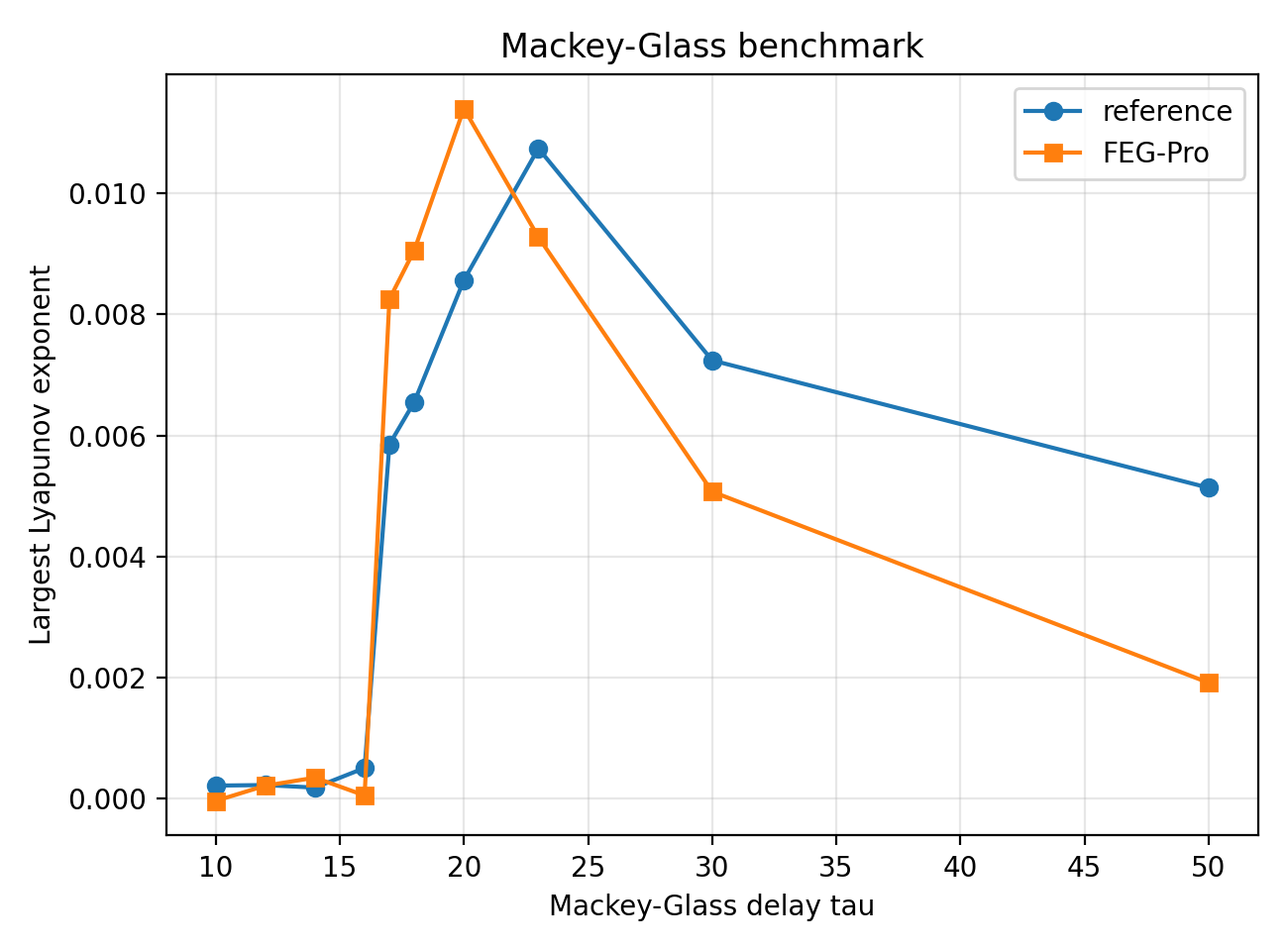}
    \caption{Mackey-Glass delay benchmark: reference largest Lyapunov exponent versus \FEGPro estimate for different delays.}
    \label{fig:mg-scatter}
\end{figure}

The Mackey-Glass benchmark tests delayed scalar dynamics. Figure~\ref{fig:mg-scatter} shows the reference-versus-estimate relation, and Table~\ref{tab:mg-detailed} gives the detailed values. The cleanest behavior occurs for intermediate cases such as \(\tau=14\) and \(\tau=18\), where the forecast-error curves are close to smooth linear growth. The \(\tau=23\) case is especially important for this manuscript: the full-record estimate is close to the reference, but the profile is classified as curved growth. It is therefore used later as the representative delayed-system case in the length-halving experiment.

\begin{table}[t]
\centering
\caption{Mackey-Glass full-record results. The table reports formal fit quantities rather than qualitative diagnostic labels.}
\label{tab:mg-detailed}
\scriptsize
\begin{tabular}{rrrrlrrr}
\toprule
\(\tau\) & Reference & Estimate & Abs. error & \code{selected\_fit} & \(R^2_{\rm single}\) & \(R^2_{\rm quad}\) & early points \\
\midrule
10 & 0.000218 & -0.000035 & 0.000253 & \code{single\_line} & 0.2569 & 0.2657 & 2 \\
12 & 0.000227 & 0.000217 & 0.000010 & \code{single\_line} & 0.9004 & 0.9075 & 2 \\
14 & 0.000185 & 0.000349 & 0.000164 & \code{single\_line} & 0.9978 & 0.9981 & 2 \\
16 & 0.000509 & 0.000055 & 0.000454 & \code{single\_line} & 0.7890 & 0.9249 & 2 \\
17 & 0.005853 & 0.008251 & 0.002398 & \code{single\_line} & 0.9981 & 0.9989 & 2 \\
18 & 0.006552 & 0.009057 & 0.002505 & \code{single\_line} & 0.9993 & 0.9995 & 2 \\
20 & 0.008566 & 0.011398 & 0.002832 & \code{single\_line} & 0.9756 & 0.9995 & 2 \\
23 & 0.010741 & 0.009276 & 0.001465 & \code{single\_line} & 0.9854 & 0.9996 & 2 \\
30 & 0.007241 & 0.005072 & 0.002169 & \code{single\_line} & 0.9830 & 0.9987 & 2 \\
50 & 0.005136 & 0.001916 & 0.003220 & \code{single\_line} & 0.9596 & 0.9778 & 2 \\
\bottomrule
\end{tabular}
\end{table}

\begin{figure}[p]
\centering
\includegraphics[width=0.95\textwidth]{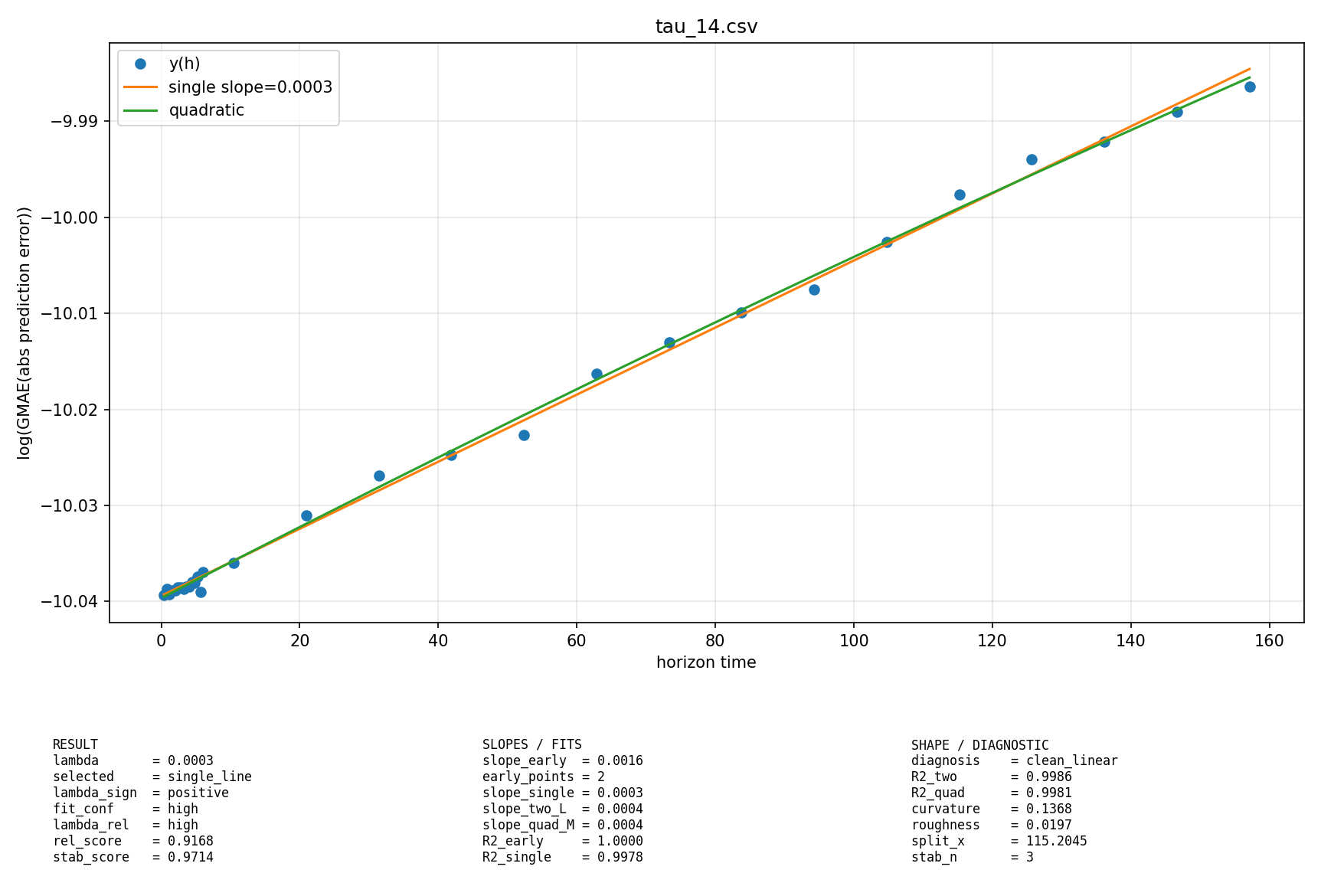}
\caption{Representative Mackey-Glass diagnostic curve for \(\tau=14\). This case illustrates a nearly linear forecast-error curve with strong single-line support.}
\label{fig:mg-tau14}
\end{figure}

\begin{figure}[p]
\centering
\includegraphics[width=0.95\textwidth]{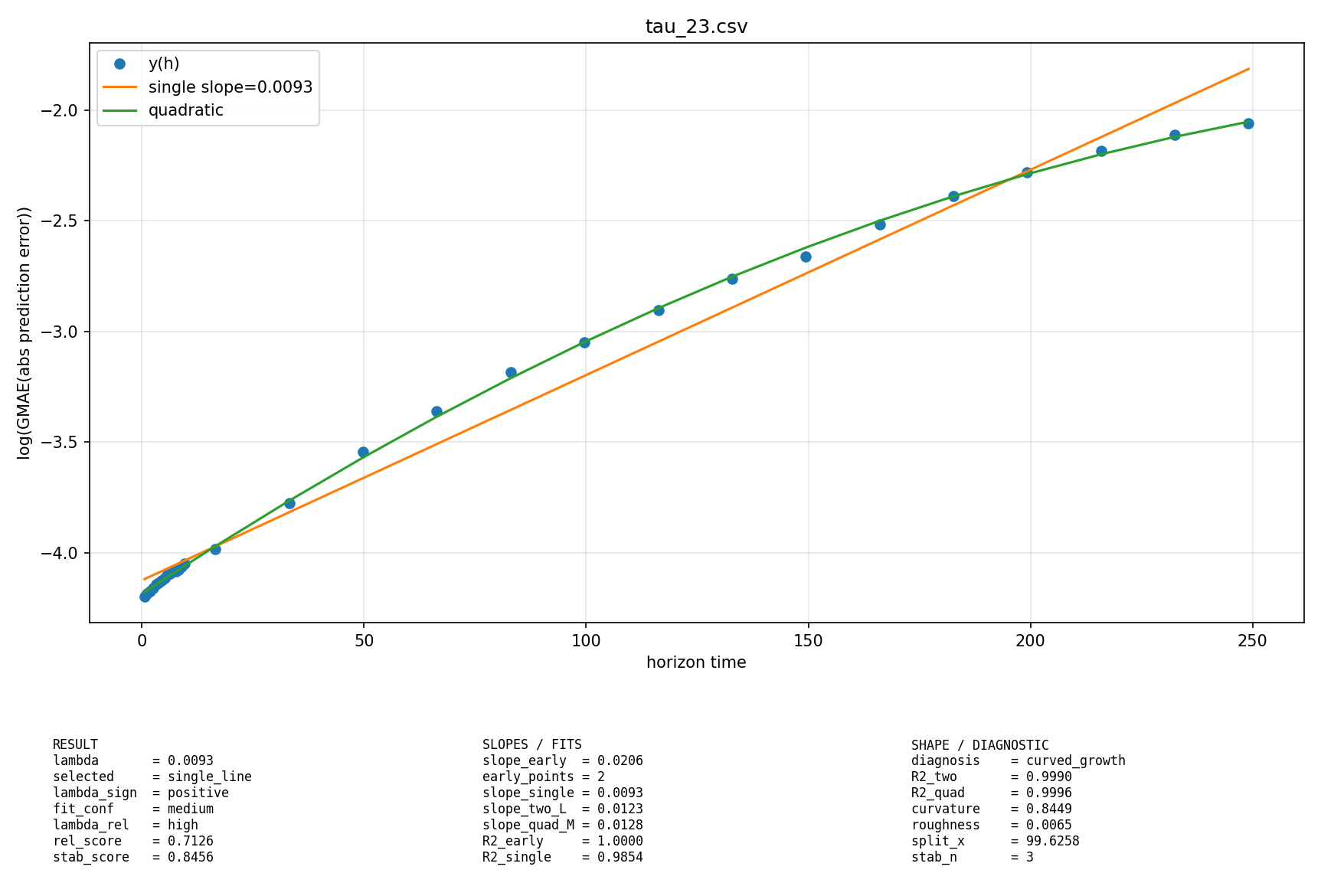}
\caption{Representative Mackey-Glass diagnostic curve for \(\tau=23\). The estimate is close to the reference value, but the curve has a smooth nonlinear component. The reported slope is therefore read together with curvature and roughness descriptors.}
\label{fig:mg-tau23}
\end{figure}

\subsubsection{Lorenz-63 scalar-observable benchmark}

\begin{figure}[t]
    \centering
    \includegraphics[width=0.82\textwidth]{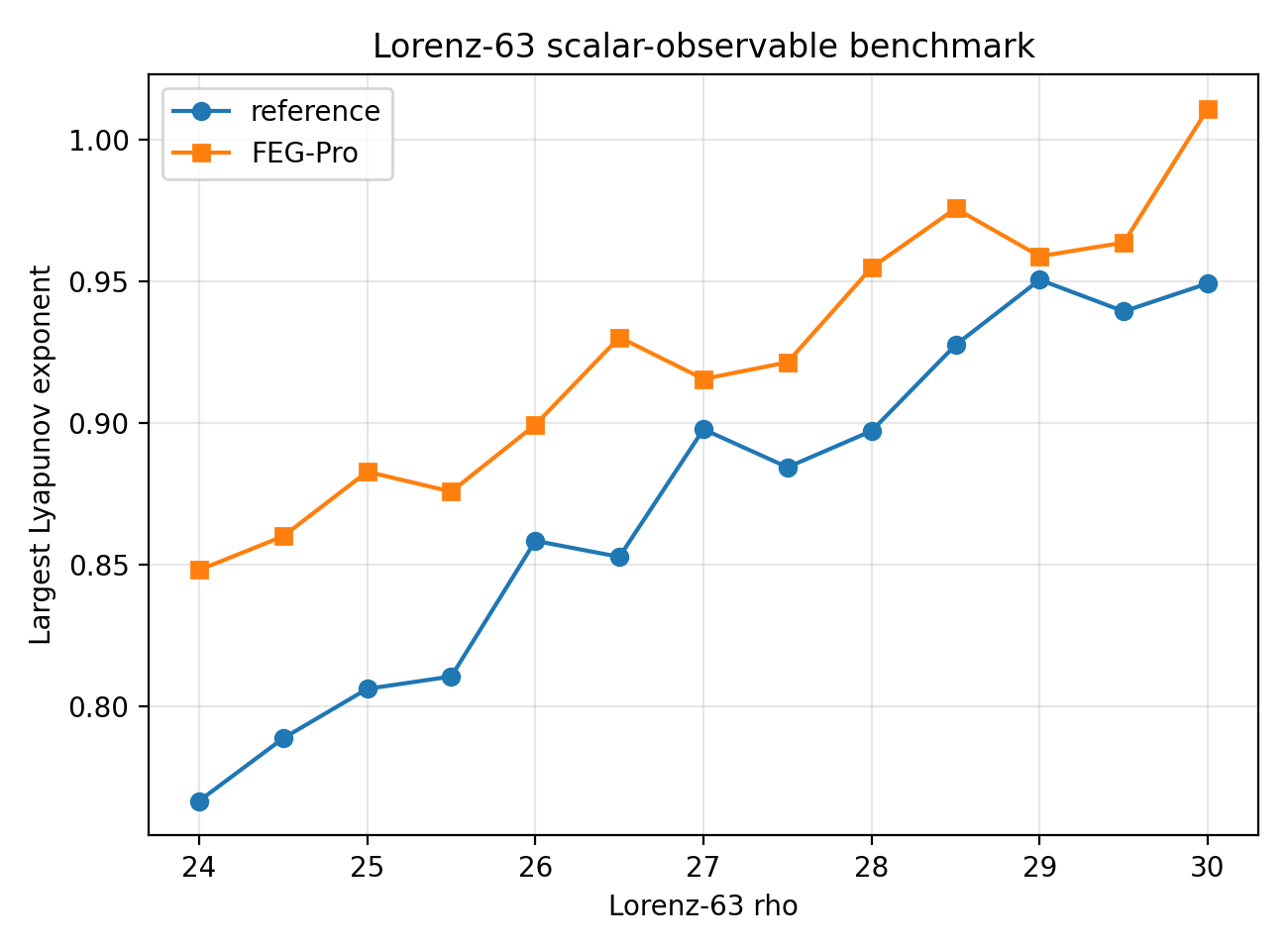}
    \caption{Lorenz-63 scalar-module benchmark: full-state variational reference exponent versus \FEGPro estimate from the scalar observable \(r(t)\).}
    \label{fig:lorenz-scatter}
\end{figure}

The Lorenz benchmark is an observable-based test rather than a full-state tangent-space calculation. \FEGPro sees only \(r(t)\), while the reference exponent is computed from the full Lorenz variational equations. Figure~\ref{fig:lorenz-scatter} shows that the scalar-observable estimate tracks the trend with \(\rho\), with a moderate upward bias. Table~\ref{tab:lorenz-detailed} shows that the forecast-error curves have very high \(R^2_{\rm single}\); in the common configuration the reported slope is the single-line regime for all Lorenz records. The \(\rho=27\) case is used later as the representative continuous-time case because its full-record estimate is close to the reference.

\begin{table}[t]
\centering
\caption{Lorenz-63 scalar-module full-record results. The table reports the formal slope-selection regime and numerical fit support.}
\label{tab:lorenz-detailed}
\scriptsize
\begin{tabular}{rrrrlrr}
\toprule
\(\rho\) & Reference & Estimate & Abs. error & \code{selected\_fit} & \(R^2_{\rm single}\) & \(R^2_{\rm quad}\) \\
\midrule
24.0 & 0.7665 & 0.8481 & 0.0816 & \code{single\_line} & 0.9999 & 0.9999 \\
24.5 & 0.7886 & 0.8600 & 0.0714 & \code{single\_line} & 0.9995 & 0.9995 \\
25.0 & 0.8061 & 0.8828 & 0.0767 & \code{single\_line} & 0.9991 & 0.9991 \\
25.5 & 0.8105 & 0.8758 & 0.0653 & \code{single\_line} & 0.9994 & 0.9996 \\
26.0 & 0.8584 & 0.8994 & 0.0410 & \code{single\_line} & 0.9997 & 0.9997 \\
26.5 & 0.8527 & 0.9303 & 0.0775 & \code{single\_line} & 0.9993 & 0.9994 \\
27.0 & 0.8978 & 0.9156 & 0.0178 & \code{single\_line} & 0.9994 & 0.9994 \\
27.5 & 0.8844 & 0.9214 & 0.0371 & \code{single\_line} & 0.9996 & 0.9997 \\
28.0 & 0.8972 & 0.9549 & 0.0577 & \code{single\_line} & 0.9992 & 0.9994 \\
28.5 & 0.9276 & 0.9758 & 0.0482 & \code{single\_line} & 0.9994 & 0.9994 \\
29.0 & 0.9506 & 0.9589 & 0.0083 & \code{single\_line} & 0.9991 & 0.9991 \\
29.5 & 0.9394 & 0.9637 & 0.0243 & \code{single\_line} & 0.9991 & 0.9993 \\
30.0 & 0.9493 & 1.0111 & 0.0617 & \code{single\_line} & 0.9993 & 0.9994 \\
\bottomrule
\end{tabular}
\end{table}

\FloatBarrier

\section{Length-halving robustness experiment}

\subsection{Design}

The full-record results establish that \FEGPro can recover useful finite-horizon slopes close to reference exponents in favorable cases and can flag less ideal cases through diagnostics. The next question is how the profile behaves when the same record is shortened. We therefore selected one representative series from each benchmark family:
\begin{itemize}
    \item logistic map \(r=4\), a memoryless discrete map with a clean theoretical exponent;
    \item Mackey-Glass \(\tau=23\), a delayed scalar system with a curved but informative full-record profile;
    \item Lorenz-63 scalar module at \(\rho=27\), a continuous-time scalar-observable case with a clean full-record profile.
\end{itemize}
Each full record was recursively split in half. At level \(j\), the record is divided into
\begin{equation}
    n_j = 2^j
\end{equation}
contiguous fragments of approximately equal length, each of nominal size
\begin{equation}
    N_j \approx \frac{N_0}{2^j},
\end{equation}
where \(N_0\) is the full-record length. \FEGPro is then applied independently to every fragment without retuning. Splitting stops when the next level would produce fragments shorter than 500 samples. In favorable cases all \(n_j\) fragments yield numerical outputs, whereas in more difficult cases some fragments are classified as unavailable. In the figures below, the annotation \(n=\cdot\) denotes the number of valid numerical fragment outputs actually contributing to the plotted mean and standard deviation at that level. Thus, for the logistic and Lorenz examples \(n\) coincides with \(2^j\), while for the shortest Mackey--Glass \(\tau=23\) level only four fragments produced valid numerical profiles.

The primary endpoint is \(\lambda_{\mathrm{FEG}}\). To test the profile idea, we also monitor selected secondary descriptors: roughness \(D_{\mathrm{rough}}\), curvature \(D_{\mathrm{curv}}\), monotonicity \(D_{\mathrm{mono}}\), and mean entropy \(\overline{\FEDE}\). The main figures focus on \(\lambda_{\mathrm{FEG}}\), \(D_{\mathrm{rough}}\), and \(\overline{\FEDE}\), because these gave the clearest interpretation in the present experiment. Open circles show individual fragment values, which makes the visual compactness or dispersion of each descriptor directly visible.

\subsection{Primary slope response}

\begin{figure}[t]
    \centering
    \includegraphics[width=\textwidth]{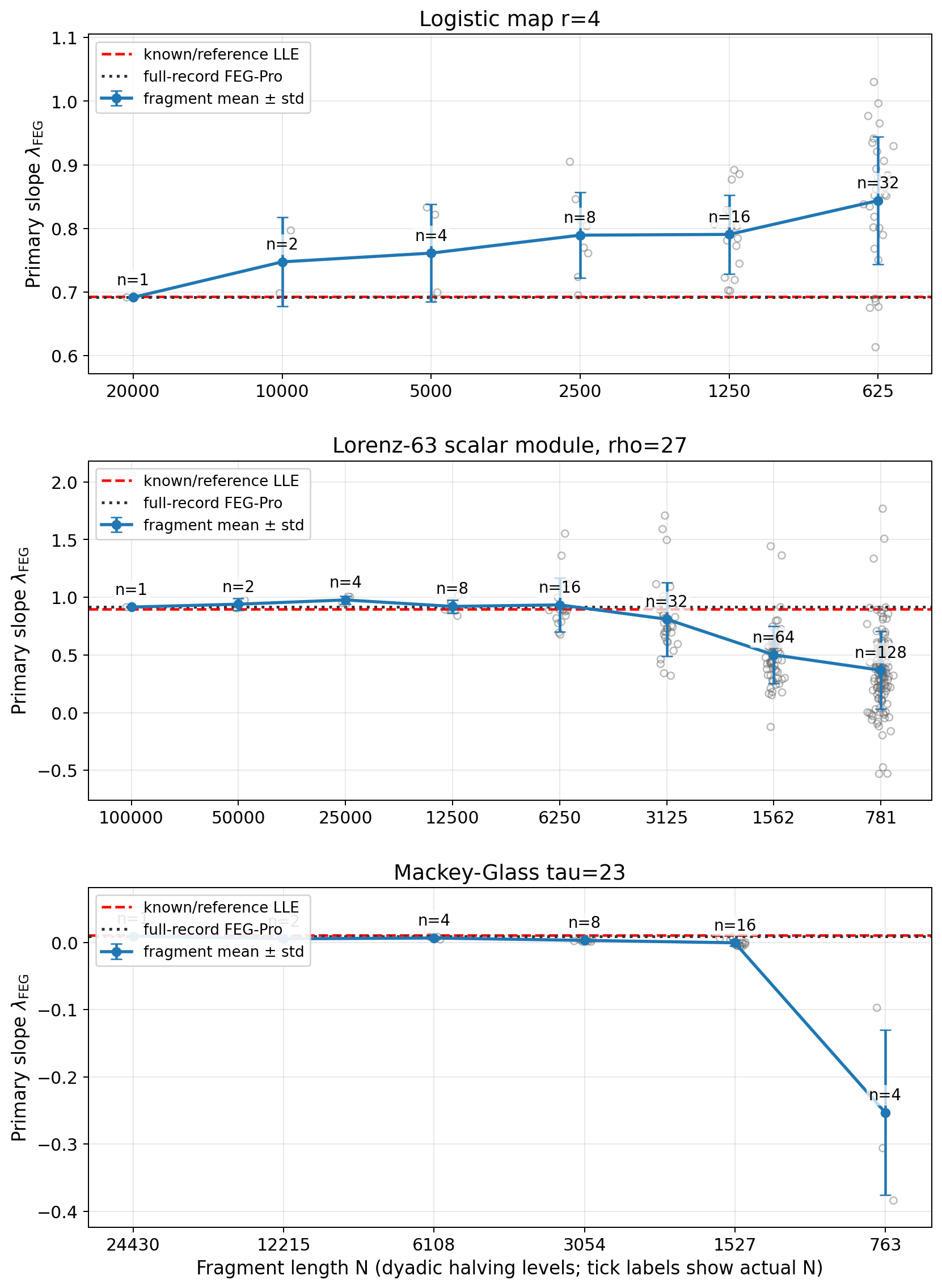}
    \caption{Primary \(\lambda_{\mathrm{FEG}}\) response under dyadic record shortening. Open circles show individual fragment estimates. Blue markers and bars show the mean and one standard deviation across valid fragments at the same length. The dotted black line is the full-record \FEGPro value, whereas the dashed red line is the known/reference Lyapunov value used in the benchmark tables. The label \(n=\cdot\) above each mean indicates the number of valid fragment estimates contributing to that summary.}
    \label{fig:length-lambda}
\end{figure}

Figure~\ref{fig:length-lambda} shows the response of the primary slope. The blue curve shows the fragment mean \(\pm\) one standard deviation, open circles show the individual fragment estimates, the dotted black line marks the full-record \(\lambda_{\mathrm{FEG}}\) value, and the dashed red line marks the known/reference Lyapunov value from the benchmark tables. Thus the figure separates two different reference levels: the ideal or externally tabulated value and the value obtained by \FEGPro on the full record. The logistic map remains usable down to the shortest tested fragments, but the mean slope becomes biased upward. The Lorenz example remains close to both reference levels through intermediate lengths and then shows a sharp increase in variability. The Mackey--Glass \(\tau=23\) example is more fragile: by \(N=1527\) the mean slope is close to zero, and at \(N=763\) only four fragments produced valid numerical slope descriptors.

\subsection{Secondary profile descriptors}

\begin{figure}[t]
    \centering
    \includegraphics[width=0.92\textwidth]{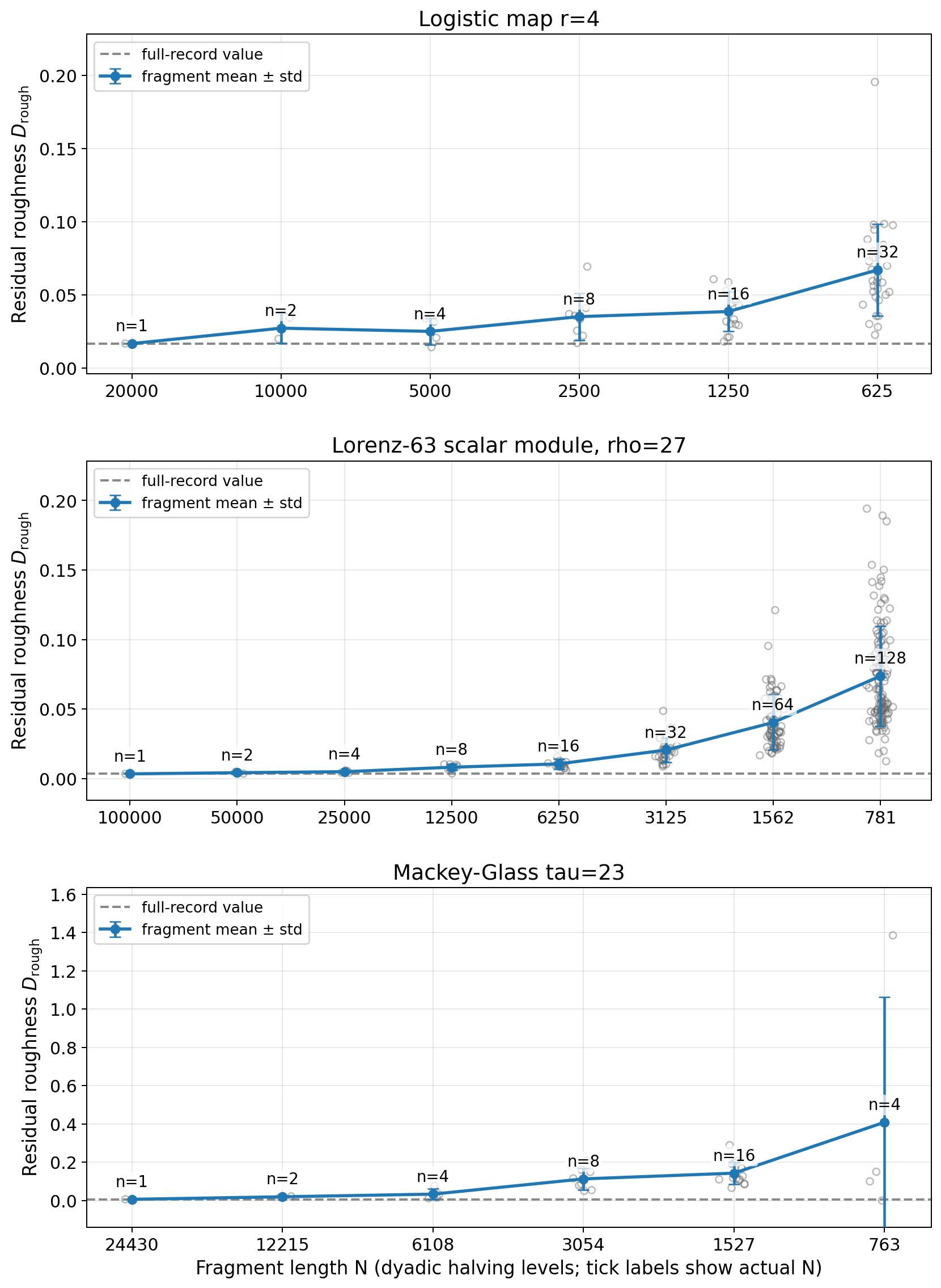}
    \caption{Residual roughness after quadratic detrending under dyadic shortening. Open circles show individual fragment values; blue markers and bars show mean \(\pm\) standard deviation over valid fragments. The descriptor increases as the available record becomes shorter in all three examples. In the Mackey--Glass \(\tau=23\) case, the shortest level has only four valid numerical profile values and should be interpreted as a failure regime.}
    \label{fig:length-roughness}
\end{figure}

\begin{figure}[t]
    \centering
    \includegraphics[width=0.92\textwidth]{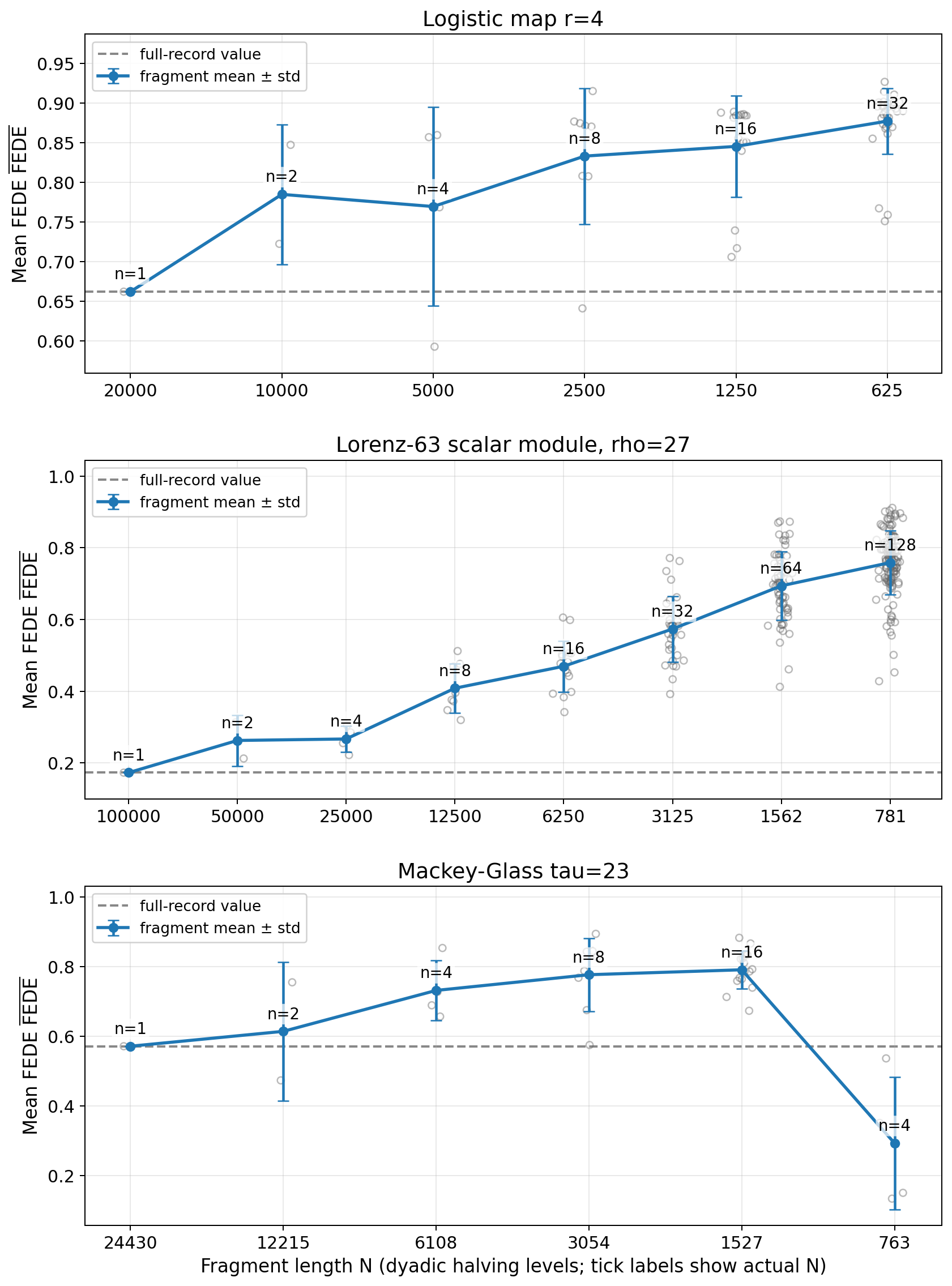}
    \caption{Mean normalized FEDE under dyadic shortening. Open circles show individual fragment values; blue markers and bars show mean \(\pm\) standard deviation over valid fragments. FEDE increases compactly in the logistic and Lorenz examples and remains informative for Mackey--Glass \(\tau=23\) down to the last pre-failure level.}
    \label{fig:length-fede}
\end{figure}

Figures~\ref{fig:length-roughness} and~\ref{fig:length-fede} show two secondary descriptors selected for the main text. The roughness descriptor increases with shortening in all three examples. For Lorenz-63, it rises from 0.00337 on the full record to 0.07362\(\pm\)0.03583 at \(N=781\). For Mackey-Glass \(\tau=23\), it grows from 0.00629 to 0.14290\(\pm\)0.05774 at \(N=1527\). The final \(N=763\) level is a failure regime: only four fragments produced valid numerical profiles.

Mean FEDE provides a complementary distributional view. In the logistic map it increases from 0.6619 to 0.8773\(\pm\)0.0416 at \(N=625\), with compact spread at the shortest length. In the Lorenz case it increases almost monotonically from 0.1727 to 0.7585\(\pm\)0.0894. Mackey-Glass \(\tau=23\) shows a useful increase up to \(N=1527\), from 0.5714 to 0.7911\(\pm\)0.0550; the shortest level is again treated as a failure regime.

\begin{table}[t]
\centering
\caption{Compact numerical summary of the length-halving experiment. Values are means \(\pm\) standard deviations over fragments.}
\label{tab:length-compact}
\small
\begin{tabular}{P{0.20\textwidth}P{0.13\textwidth}P{0.21\textwidth}P{0.21\textwidth}P{0.18\textwidth}}
\toprule
Series & Length range & \(\lambda_{\mathrm{FEG}}\) & \(D_{\rm rough}\) & \(\overline{\FEDE}\) \\
\midrule
Logistic map $r=4$ & 20000 $\rightarrow$ 625 & 0.6916 $\rightarrow$ 0.8438$\pm$0.1005 & 0.01661 $\rightarrow$ 0.06693$\pm$0.03155 & 0.6619 $\rightarrow$ 0.8773$\pm$0.0416 \\
Lorenz-63 module $\rho=27$ & 100000 $\rightarrow$ 781 & 0.9156 $\rightarrow$ 0.3680$\pm$0.3374 & 0.00337 $\rightarrow$ 0.07362$\pm$0.03583 & 0.1727 $\rightarrow$ 0.7585$\pm$0.0894 \\
Mackey-Glass $\tau=23$ & 24430 $\rightarrow$ 1527 & 0.0093 $\rightarrow$ -0.0004$\pm$0.0047; at $N=763$, only 4 valid slope values & 0.00629 $\rightarrow$ 0.14290$\pm$0.05774 & 0.5714 $\rightarrow$ 0.7911$\pm$0.0550 \\
\bottomrule
\end{tabular}
\end{table}

The key observation is that shortening does not simply add unstructured noise to every output. Some secondary descriptors, especially residual roughness and mean FEDE, show structured and often monotone responses. They therefore act as indicators of profile degradation. This is the central reason for treating them as part of the method rather than as discarded implementation diagnostics.

\FloatBarrier

\section{Discussion}

The full-record benchmarks show that \FEGPro is useful but should be interpreted carefully. In favorable discrete-map cases, such as the logistic and skew-tent maps, \(\lambda_{\mathrm{FEG}}\) is close to the reference exponent. In Lorenz-63 scalar observables, the method captures the correct magnitude and trend, although the scalar-module estimate is moderately biased relative to the full-state variational exponent. In Mackey-Glass delay dynamics, several cases are recovered well, while curved or weak profiles are identified by numerical fit and shape quantities such as \(R^2_{\rm single}\), \(R^2_{\rm quad}\), and the roughness descriptors.

This behavior supports the main methodological claim: the forecast-error curve itself is the primary object. The slope is important, but it is only defensible when the profile has the right geometry. A large or small slope without diagnostics can be misleading. A slope reported together with the formal fit-selection regime, linearity, curvature, roughness, monotonicity, and FEDE is more informative, because it states not only the estimated finite-horizon instability but also the shape evidence behind it.

A positive value of \(\lambda_{\mathrm{FEG}}\) should not be taken by itself as a proof of chaos. It indicates finite-horizon growth of forecast errors for the chosen observable, reconstruction, predictor, horizon grid, and fitting rule. In deterministic benchmark systems with a supported quasi-linear forecast-error profile, this growth can be consistent with the largest Lyapunov exponent. In experimental, noisy, nonstationary, or poorly reconstructed signals, however, positive forecast-error growth may also arise from model mismatch, measurement noise, finite-sample effects, or drift. The sign and magnitude of \(\lambda_{\mathrm{FEG}}\) should therefore be interpreted jointly with the profile diagnostics.

The length-halving experiment strengthens this interpretation. The primary slope responds differently across the three representative systems: the logistic map shows an upward bias under shortening, Lorenz-63 remains stable through intermediate lengths before degrading, and Mackey-Glass \(\tau=23\) loses stable slope support earlier. A scalar-only analysis would stop at this conclusion. The profile analysis shows more: residual roughness rises as the curve becomes locally less smooth, and FEDE rises as the signed forecast-error distribution becomes more uncertain. In the Lorenz and logistic examples these changes are compact and interpretable. In Mackey-Glass, the same descriptors reveal the transition toward profile failure.

The secondary descriptors therefore have two roles. First, they are quality indicators for the finite-horizon FEG slope. Second, they are potential features in their own right. Residual roughness measures local irregularity after removing a smooth quadratic trend. Mean FEDE measures distributional uncertainty rather than error amplitude. Monotonicity measures whether the error-growth curve preserves an increasing structure. Curvature measures the departure from a single-line representation. These quantities may be valuable in applications where the exact asymptotic exponent is not available or not the only relevant property.

This feature-oriented role is a central practical motivation for the method. In many applied signal-processing problems, especially those addressed by machine learning, the goal is not only to estimate an invariant of a known system but to extract robust, interpretable coordinates of the observed signal. The results here suggest that forecast-error shape and distribution descriptors can carry information complementary to \(\lambda_{\mathrm{FEG}}\). Their monotone or compact response under record shortening is only an indirect validation, not a complete application study, but it supports the hypothesis that these descriptors can serve as useful features for future classification, clustering, change detection, or comparative analysis of nonlinear signals.

The present version deliberately focuses on length robustness rather than adding a large noise experiment. Additive noise will predictably affect errors, roughness, and entropy; a rigorous noise study would require its own calibration and comparison protocol. Similarly, a preliminary off-the-shelf Rosenstein-style baseline was kept in the computational archive but not presented as a formal comparator. In the logistic-map case it underestimates the expected exponent without additional tuning of embedding and fitting settings. A fair comparison with classical algorithms should therefore be treated as a separate methodological study.

Several limitations remain. The current configuration is universal but not guaranteed optimal. The kNN forecast model is simple, and other local predictors could be tested. The FEDE histogram uses a fixed bin count, which is convenient but may not be optimal for all error distributions. The method also produces finite-horizon, observable-based estimates; these should not be confused with exact tangent-space exponents when the full dynamical system is known.

The deterministic fit-selection rule should be understood as an empirical operational criterion. It is reproducible and was chosen to behave consistently across the present benchmark families, but it is not a theoretically unique definition of a Lyapunov exponent. A systematic comparison with classical LLE algorithms, including careful retuning of embedding delays, dimensions, Theiler windows, neighborhood sizes, and fitting intervals, is outside the scope of the present manuscript. Real-world biomedical or experimental applications are also left for separate studies, where the profile descriptors can be evaluated as downstream features rather than as benchmark estimates against known exponents. These limitations are intentional in the present manuscript: the goal is to establish the FEG-Pro profile and its operational diagnostics before optimizing every component. Future work can replace the local predictor, tune the entropy estimator, or compare against classical algorithms without changing the central object of analysis, namely the multi-horizon forecast-error growth profile.

\section{Conclusion}

We introduced \FEGPro, a forecast-error growth profiling pipeline for nonlinear scalar time series. The method generates a finite-horizon forecast-error growth slope using a deterministic fit-selection rule, and it also generates interpretable descriptors of curve shape, residual roughness, monotonicity, and forecast-error distribution entropy. Expanded benchmark definitions and full-record results show where the method performs well and where the profile warns against over-interpreting a single slope. A dyadic length-halving experiment shows that secondary descriptors, especially residual roughness and mean FEDE, can behave monotonically and remain informative as the available record shortens.

The main practical implication is that the forecast-error curve can be used as a source of features, not only as a path to a single exponent-related number. In future studies, the proposed descriptors can be supplied to machine-learning or statistical signal-analysis pipelines as compact, physically motivated features of nonlinear predictability. They may be useful when one needs to compare signals, detect degradation, characterize finite-data profile stability, or classify regimes in settings where a reference exponent is unknown. In this sense, \(\lambda_{\mathrm{FEG}}\) remains the primary endpoint, but it is embedded in a richer feature vector that describes how the prediction errors grow, how smooth or curved the growth curve is, and how the signed error distribution changes with horizon.

The present work should therefore be read as a methodological step from scalar exponent estimation toward forecast-error growth profiling. The benchmark and length-halving experiments provide evidence that the additional descriptors are meaningful and interpretable, but their full value as features will require dedicated downstream studies on experimental and domain-specific signals. This is also where the framework can be extended: alternative local predictors, adaptive entropy estimation, additional profile descriptors, and systematic comparisons with classical LLE algorithms can all be incorporated without changing the core idea that the multi-horizon forecast-error curve is the object of analysis.

\clearpage
\section*{CRediT authorship contribution statement}
Andrei Velichko: Conceptualization, Methodology, Software, Formal analysis, Investigation, Data curation, Visualization, Writing - original draft, Writing - review and editing, Project administration, Funding acquisition. N'Gbo N'Gbo: Validation, Investigation, Writing - review and editing. Bruno Carpentieri: Methodology, Supervision, Writing - review and editing. Mudassir Shams: Methodology, Validation, Writing - review and editing.

\section*{Declaration of competing interest}
The authors declare no competing interests.

\section*{Data and code availability}
The benchmark data tables, representative forecast-error plots, and dyadic length-halving summaries used in this draft are included in the project archive. The archive contains the full-record benchmark tables and the length-halving summary tables in CSV format. A public code repository or preprint package will be prepared for the final version of the work.

\section*{Funding}
This research was supported by the Russian Science Foundation (grant no. 22-11-00055-P, \url{https://rscf.ru/en/project/22-11-00055/}, accessed on 10 June 2025).

\section*{Acknowledgements}
The authors thank colleagues involved in earlier discussions of forecast-error-based Lyapunov estimation and nonlinear time-series diagnostics.

\section*{Declaration of generative AI and AI-assisted technologies in the manuscript preparation process}

During the preparation of this work, the authors used OpenAI ChatGPT to assist with English-language editing, translation of author-prepared draft text into English, improvement of clarity and style, and technical polishing of the LaTeX manuscript. After using this tool, the authors reviewed and edited the content as needed and take full responsibility for the content of the published article.

\bibliographystyle{elsarticle-num}
\bibliography{references}

@article{shannon1948,
  author = {Shannon, Claude E.},
  title = {A Mathematical Theory of Communication},
  journal = {The Bell System Technical Journal},
  year = {1948},
  volume = {27},
  number = {3},
  pages = {379--423}
}

@article{lorenz1963,
  author = {Lorenz, Edward N.},
  title = {Deterministic Nonperiodic Flow},
  journal = {Journal of the Atmospheric Sciences},
  year = {1963},
  volume = {20},
  number = {2},
  pages = {130--141}
}

@article{mackey1977,
  author = {Mackey, Michael C. and Glass, Leon},
  title = {Oscillation and Chaos in Physiological Control Systems},
  journal = {Science},
  year = {1977},
  volume = {197},
  number = {4300},
  pages = {287--289}
}

@article{takens1981,
  author = {Takens, Floris},
  title = {Detecting Strange Attractors in Turbulence},
  journal = {Lecture Notes in Mathematics},
  year = {1981},
  volume = {898},
  pages = {366--381}
}

@article{wolf1985,
  author = {Wolf, Alan and Swift, Jack B. and Swinney, Harry L. and Vastano, John A.},
  title = {Determining Lyapunov Exponents from a Time Series},
  journal = {Physica D: Nonlinear Phenomena},
  year = {1985},
  volume = {16},
  number = {3},
  pages = {285--317}
}

@article{fraser1986,
  author = {Fraser, Andrew M. and Swinney, Harry L.},
  title = {Independent Coordinates for Strange Attractors from Mutual Information},
  journal = {Physical Review A},
  year = {1986},
  volume = {33},
  number = {2},
  pages = {1134--1140}
}

@article{kennel1992,
  author = {Kennel, Matthew B. and Brown, Reggie and Abarbanel, Henry D. I.},
  title = {Determining Embedding Dimension for Phase-space Reconstruction Using a Geometrical Construction},
  journal = {Physical Review A},
  year = {1992},
  volume = {45},
  number = {6},
  pages = {3403--3411}
}

@article{rosenstein1993,
  author = {Rosenstein, Michael T. and Collins, James J. and De Luca, Carlo J.},
  title = {A Practical Method for Calculating Largest Lyapunov Exponents from Small Data Sets},
  journal = {Physica D: Nonlinear Phenomena},
  year = {1993},
  volume = {65},
  number = {1--2},
  pages = {117--134}
}

@article{kantz1994,
  author = {Kantz, Holger},
  title = {A Robust Method to Estimate the Maximal Lyapunov Exponent of a Time Series},
  journal = {Physics Letters A},
  year = {1994},
  volume = {185},
  number = {1},
  pages = {77--87}
}

@book{abarbanel1996,
  author = {Abarbanel, Henry D. I.},
  title = {Analysis of Observed Chaotic Data},
  publisher = {Springer},
  year = {1996}
}

@article{parlitz2016,
  author = {Parlitz, Ulrich},
  title = {Estimating Lyapunov Exponents from Time Series},
  journal = {Chaos Detection and Predictability},
  year = {2016},
  pages = {1--34},
  doi = {10.1007/978-3-662-48410-4_1}
}

@article{ma2006,
  author = {Ma, Hongguang and Han, Chong-zhao},
  title = {Selection of Embedding Dimension and Delay Time in Phase Space Reconstruction},
  journal = {Frontiers of Electrical and Electronic Engineering in China},
  year = {2006},
  volume = {1},
  pages = {111--114},
  doi = {10.1007/s11460-005-0023-7}
}

@article{matilla2021,
  author = {Matilla-García, Mariano and Morales, Isabel and Rodríguez, Javier and Marín, Manuel R.},
  title = {Selection of Embedding Dimension and Delay Time in Phase Space Reconstruction via Symbolic Dynamics},
  journal = {Entropy},
  year = {2021},
  volume = {23},
  number = {2},
  pages = {221},
  doi = {10.3390/e23020221}
}

@article{tamma2016,
  author = {Tamma, Aniruddha and Khubchandani, Bhushan},
  title = {Accurate Determination of Time Delay and Embedding Dimension for State Space Reconstruction from a Scalar Time Series},
  journal = {arXiv: Chaotic Dynamics},
  year = {2016}
}

@article{zhu2016,
  author = {Zhu, Shengli and Gan, Lu},
  title = {Incomplete Phase-space Method to Reveal Time Delay from Scalar Time Series},
  journal = {Physical Review E},
  year = {2016},
  volume = {94},
  number = {5},
  pages = {052210},
  doi = {10.1103/physreve.94.052210}
}

@article{zeng1991,
  author = {Zeng, Xubin and Eykholt, Richard and Pielke, Roger A.},
  title = {Estimating the Lyapunov-exponent Spectrum from Short Time Series of Low Precision},
  journal = {Physical Review Letters},
  year = {1991},
  volume = {66},
  number = {25},
  pages = {3229--3232},
  doi = {10.1103/physrevlett.66.3229}
}

@article{liu2005,
  author = {Liu, Hai-feng and Dai, Zhenghua and Li, Wei-feng and Gong, Xin and Yu, Zun-hong},
  title = {Noise Robust Estimates of the Largest Lyapunov Exponent},
  journal = {Physics Letters A},
  year = {2005},
  volume = {341},
  pages = {119--127},
  doi = {10.1016/j.physleta.2005.04.048}
}

@article{yao2012,
  author = {Yao, Tian-Liang and Liu, Hai-feng and Xu, Jian-Liang and Li, Wei-feng},
  title = {Estimating the Largest Lyapunov Exponent and Noise Level from Chaotic Time Series},
  journal = {Chaos},
  year = {2012},
  volume = {22},
  number = {3},
  pages = {033102},
  doi = {10.1063/1.4731800}
}

@article{mehdizadeh2017,
  author = {Mehdizadeh, Sina and Sanjari, M. A.},
  title = {Effect of Noise and Filtering on Largest Lyapunov Exponent of Time Series Associated with Human Walking},
  journal = {Journal of Biomechanics},
  year = {2017},
  volume = {64},
  pages = {236--239},
  doi = {10.1016/j.jbiomech.2017.09.009}
}

@article{mehdizadeh2018,
  author = {Mehdizadeh, Sina},
  title = {A Robust Method to Estimate the Largest Lyapunov Exponent of Noisy Signals: A Revision to the Rosenstein's Algorithm},
  journal = {bioRxiv},
  year = {2018},
  doi = {10.1101/381111}
}

@article{escot2023,
  author = {Escot, L. and Sandubete, Julio E.},
  title = {Estimating Lyapunov Exponents on a Noisy Environment by Global and Local Jacobian Indirect Algorithms},
  journal = {Applied Mathematics and Computation},
  year = {2023},
  volume = {436},
  pages = {127498},
  doi = {10.1016/j.amc.2022.127498}
}

@article{farmer1987,
  author = {Farmer, J. Doyne and Sidorowich, John J.},
  title = {Predicting Chaotic Time Series},
  journal = {Physical Review Letters},
  year = {1987},
  volume = {59},
  number = {8},
  pages = {845--848},
  doi = {10.1103/physrevlett.59.845}
}

@article{casdagli1989,
  author = {Casdagli, Martin},
  title = {Nonlinear Prediction of Chaotic Time Series},
  journal = {Physica D: Nonlinear Phenomena},
  year = {1989},
  volume = {35},
  pages = {335--356},
  doi = {10.1016/0167-2789(89)90074-2}
}

@article{sugihara1990,
  author = {Sugihara, George and May, Robert M.},
  title = {Nonlinear Forecasting as a Way of Distinguishing Chaos from Measurement Error in Time Series},
  journal = {Nature},
  year = {1990},
  volume = {344},
  pages = {734--741},
  doi = {10.1038/344734a0}
}

@article{sugihara1994,
  author = {Sugihara, George},
  title = {Nonlinear Forecasting for the Classification of Natural Time Series},
  journal = {Philosophical Transactions of the Royal Society of London. Series A},
  year = {1994},
  volume = {348},
  pages = {477--495},
  doi = {10.1098/rsta.1994.0106}
}

@article{hegger1999,
  author = {Hegger, Rainer and Kantz, Holger and Schreiber, Thomas},
  title = {Practical Implementation of Nonlinear Time Series Methods: The TISEAN Package},
  journal = {Chaos},
  year = {1999},
  volume = {9},
  number = {2},
  pages = {413--435},
  doi = {10.1063/1.166424}
}

@book{kantzschreiber2004,
  author = {Kantz, Holger and Schreiber, Thomas},
  title = {Nonlinear Time Series Analysis},
  publisher = {Cambridge University Press},
  year = {2004},
  edition = {2}
}

@article{chang2017,
  author = {Chang, Chun-Wei and Ushio, Masayuki and Hsieh, Chih-hao},
  title = {Empirical Dynamic Modeling for Beginners},
  journal = {Ecological Research},
  year = {2017},
  volume = {32},
  pages = {785--796},
  doi = {10.1007/s11284-017-1469-9}
}

@article{bradley2015,
  author = {Bradley, Elizabeth and Kantz, Holger},
  title = {Nonlinear Time-series Analysis Revisited},
  journal = {Chaos},
  year = {2015},
  volume = {25},
  number = {9},
  pages = {097610},
  doi = {10.1063/1.4917289}
}

@article{fulcher2017,
  author = {Fulcher, Ben D.},
  title = {Feature-Based Time-Series Analysis},
  journal = {Feature Engineering for Machine Learning and Data Analytics},
  year = {2017},
  pages = {87--116},
  doi = {10.1201/9781315181080-4}
}

@article{lubba2019,
  author = {Lubba, Carl H. and Sethi, Sarab S. and Knaute, Philip and Schultz, Simon R. and Fulcher, Ben D. and Jones, Nick S.},
  title = {catch22: CAnonical Time-series CHaracteristics},
  journal = {Data Mining and Knowledge Discovery},
  year = {2019},
  volume = {33},
  pages = {1821--1852},
  doi = {10.1007/s10618-019-00647-x}
}

@article{christ2018,
  author = {Christ, Maximilian and Braun, Nils and Neuffer, Julius and Kempa-Liehr, Andreas W.},
  title = {Time Series FeatuRe Extraction on Basis of Scalable Hypothesis Tests (tsfresh - A Python Package)},
  journal = {Neurocomputing},
  year = {2018},
  volume = {307},
  pages = {72--77},
  doi = {10.1016/j.neucom.2018.03.067}
}

@article{henderson2022,
  author = {Henderson, Trent and Fulcher, Ben D.},
  title = {Feature-Based Time-Series Analysis in R Using the theft Package},
  journal = {arXiv},
  year = {2022},
  volume = {abs/2208.06146},
  doi = {10.48550/arxiv.2208.06146}
}

@article{deoliveira2021,
  author = {de Oliveira, J. F. L. and Silva, E. and de Mattos Neto, Paulo S. G.},
  title = {A Hybrid System Based on Dynamic Selection for Time Series Forecasting},
  journal = {IEEE Transactions on Neural Networks and Learning Systems},
  year = {2021},
  volume = {33},
  pages = {3251--3263},
  doi = {10.1109/tnnls.2021.3051384}
}

@article{ilhan2024,
  author = {Ilhan, Emirhan and Koc, Ahmet B. and Kozat, Suleyman S.},
  title = {Exploiting Residual Errors in Nonlinear Online Prediction},
  journal = {Machine Learning},
  year = {2024},
  volume = {113},
  pages = {6065--6091},
  doi = {10.1007/s10994-024-06554-7}
}

@article{jensen2022,
  author = {Jensen, Vilde and Bianchi, Filippo and Anfinsen, Stian N.},
  title = {Ensemble Conformalized Quantile Regression for Probabilistic Time Series Forecasting},
  journal = {IEEE Transactions on Neural Networks and Learning Systems},
  year = {2022},
  volume = {35},
  pages = {9014--9025},
  doi = {10.1109/tnnls.2022.3217694}
}

@article{pessoa2025,
  author = {Pessoa, Pedro and Campitelli, P. and Shepherd, Douglas P. and Ozkan, Banu and Pressé, Steve},
  title = {Mamba Time Series Forecasting with Uncertainty Quantification},
  journal = {Machine Learning},
  year = {2025},
  volume = {6},
  doi = {10.1088/2632-2153/adec3b}
}

@article{guntu2020,
  author = {Guntu, R. and Yeditha, Pavan Kumar and Rathinasamy, M. and Perc, Matjaz and Marwan, Norbert and Kurths, Jürgen and Agarwal, A.},
  title = {Wavelet Entropy-based Evaluation of Intrinsic Predictability of Time Series},
  journal = {Chaos},
  year = {2020},
  volume = {30},
  number = {3},
  pages = {033117},
  doi = {10.1063/1.5145005}
}

@article{papacharalampous2021,
  author = {Papacharalampous, Georgia A. and Tyralis, Hristos and Pechlivanidis, I. and Grimaldi, S. and Volpi, E.},
  title = {Massive Feature Extraction for Explaining and Foretelling Hydroclimatic Time Series Forecastability at the Global Scale},
  journal = {Geoscience Frontiers},
  year = {2022},
  volume = {13},
  pages = {101349},
  doi = {10.1016/j.gsf.2022.101349}
}

@article{velichko2025chaos,
  author = {Velichko, A. and Belyaev, M. and Boriskov, P.},
  title = {A Novel Approach for Estimating Largest Lyapunov Exponents in One-Dimensional Chaotic Time Series Using Machine Learning},
  journal = {Chaos},
  year = {2025},
  volume = {35},
  number = {10},
  doi = {10.1063/5.0289352}
}

@inproceedings{belyaev2025esn,
  author = {Belyaev, M. and Velichko, Andrei and Putrolaynen, V.},
  title = {Forecasting Chaotic Time Series Using Echo State Networks: Limits of Predictability and Relationship with Entropy},
  booktitle = {Proceedings of SPIE},
  year = {2025},
  volume = {13803},
  pages = {138031C},
  doi = {10.1117/12.3078206}
}

@article{muruganantham2025local,
  author = {Muruganantham, Yazhini and Velichko, Andrei and Rajendran, S.},
  title = {Local Lyapunov Analysis via Micro-Ensembles: Finite-time Lyapunov Exponent Estimation and KNN-Based Predictive Comparison in Complex-Valued BAM Neural Networks},
  year = {2025}
}

@article{shams2026,
  author = {Shams, Mudassir and Velichko, Andrei and Carpentieri, Bruno},
  title = {Optimizing Parallel Schemes with Lyapunov Exponents and kNN-LLE Estimation},
  journal = {arXiv},
  year = {2026},
  volume = {abs/2601.13604},
  doi = {10.48550/arxiv.2601.13604}
}

\end{document}